%% file: main_v2.tex
\documentclass[twocolumn]{aastex631}
\usepackage{graphicx} 

\definecolor{red}{rgb}{0.75, 0.05, 0.05}
\newcommand{\hl}[1]{{{#1}}}
\usepackage{tikz} 
\PassOptionsToPackage{hyphens}{url}
\usepackage{hyperref}
\usepackage{xurl}

\begin{document}

\title{No Measurable Changes in Radio and X-ray Emission Surrounding Glitches in the Young Pulsar PSR$~$J2229+6114}
\input{auth}
\correspondingauthor{Wenke Xia}
\email{wenke.xia@mail.mcgill.ca}

\begin{abstract}
    We present our first result from an ongoing pulsar glitch monitoring campaign at the Canadian Hydrogen Intensity Mapping Experiment (CHIME), in which we analyzed the radio and X-ray emission surrounding four glitches in PSR~J2229+6114. Using daily CHIME observations, we detected a glitch in PSR~J2229+6114 in near-real time and triggered an X-ray follow-up with the Nuclear Spectroscopic Telescope Array (NuSTAR) two days after the glitch. We identified three additional glitch events in archival CHIME/Pulsar observations that coincided with an independent X-ray observing campaign with the Neutron star Interior Composition Explorer (NICER).. Our data show there is no measurable change in the source’s X-ray and radio emission during the four glitch events, in stark contrast to the post-glitch activity in high-magnetic-field, rotation-powered pulsars (RPPs), which have been observed to exhibit magnetar-like X-ray outbursts immediately after large glitches. Those high-magnetic-field (high-$B$) RPPs are considered transitional objects between ordinary RPPs and magnetars, thereby leading to a unifying neutron star model in which the inferred dipolar surface magnetic field strength serves as a unifying parameter. However, such a model remains challenged, in part, by the lack of constraints near the low-$B$ end of the high-$B$ regime, and our result provides additional evidence that magnetar-like post-glitch activity is likely more common among high-$B$ RPPs. 
\end{abstract}

\vspace{0.5em}
\section{Introduction} \label{sec:introduction}

Pulsars and magnetars are typically classified as two distinct types of neutron stars, largely due to their different emission properties and activities. Conventional rotation-powered pulsars (RPPs) are generally radio-loud, exhibiting stable thermal X-ray emission from surface cooling. Magnetars, on the other hand, are not commonly visible in the radio band and are primarily powered by their decaying internal magnetic fields, and they exhibit more frequent outbursts and variations in their X-ray emission. Magnetars also differ in their spin properties. Assuming magnetic dipole braking, magnetars generally show surface dipolar magnetic field strengths (hereafter ``$B$-field'') of the order of $B_{\rm dip}\approx10^{14} - 10^{15}$\,G\footnote{The surface dipolar $B$-field is given as $B_{\rm dip} = 3.2 \times 10^{19}\,{\rm G} \sqrt{P \dot{P}}$, for the pulse period $P$ (in seconds), and the spindown rate $\dot{P}$ (unitless). Magnetars can exhibit larger, more localized multipole components that may significantly exceed the inferred $B_{\rm dip}$ value. }, much higher than typical RPPs that usually have $B_{\rm dip} \lesssim 10^{14}$\,G. Both classes exhibit rotational glitches, but post-glitch activity in magnetars is often more dramatic, accompanied by X-ray outbursts or pulse profile variations \citep[see][for a review]{kb17}. 

Despite the apparent differences in properties of pulsars and magnetars, this dichotomous classification is challenged by the transient magnetar-like X-ray bursting behavior in two high-magnetic-field ($B_{\rm dip}\approx10^{13} - 10^{14}$\,G; hereafter ``high-$B$'') RPPs, PSRs$~$J1846$-$0258 \citep{J1846, J1846_2020} and J1119$-$6127 \citep{J1119}, immediately after large spin-up glitch events ($\Delta\nu/\nu > 10^{-6}$). These discoveries argue that high-$B$ RPPs in the potentially magnetar-strength inferred $B$-field regime (from $10^{13}$ to $10^{14}$\,G; hereafter ``high-$B$ RPP regime'') are also transitional objects in the neutron star population that bridge the emission and bursting behaviors of pulsars and magnetars. They suggest a unified neutron star model in which the inferred $B$-field serves as a unifying parameter in bursting behavior across the phenomenological gap between RPPs and magnetars \citep{kas03, pp+11}. In addition to radiative changes at X-ray energies, PSR~J1119$-$6127 also exhibited radio profile variations after a large glitch \citep{wje+11}. Despite similar sudden profile changes being observed in lower $B$-field RPPs through various mechanisms, those changes are believed to be different phenomena that are unrelated to glitch events (e.g., PSRs~J1643$-$1224 of \citealt{bkk+18} and J1713$+$0747 of \citealt{jcc+24}). 

While this unifying picture is consistent with existing evidence, a more robust confirmation still requires consideration of the behavior of RPPs near the lower end of the high-$B$ RPP regime (from $10^{12}$ to $10^{13}$\,G). To fill this gap, we have developed a daily-cadence pulsar monitoring campaign by leveraging the capability of daily all-sky revisits at the Canadian Hydrogen Intensity Mapping Experiment (CHIME) telescope, using its CHAMPSS \citep{champss_overview} and CHIME/Pulsar \citep{chimepsr_overview} digital instruments. Such a dedicated high-cadence monitoring campaign is necessary to robustly test the unifying model near the lower end of the high-$B$ RPP regime for two primary reasons. First, magnetar-like X-ray emission and bursts from post-glitch RPPs decay on timescales of weeks to months \citep[e.g., two months for PSR~J1846$-$0258;][]{J1846}; follow-up observations are required immediately after the glitch (usually within days). On the other hand, the glitch-induced radiative change in RPPs may be faint and therefore not detectable by existing all-sky X-ray monitors.  

On 2026 March 9, a glitch in the RPP PSR~J2229+6114 was detected by our near-real-time glitch alert system that we built for our monitoring campaign. PSR$~$J2229$+$6114 is an energetic young ``Vela-like" RPP with a spin period of 51.6\,ms \citep{J2229_discovery}. It has a spin-inferred surface dipolar $B$-field of $2.0\times 10^{12}\,{\rm G}$ \citep{psrcat}, placing it immediately below the high-$B$ RPP regime. We thereby triggered an X-ray follow-up through our target-of-opportunity (ToO) observation with the Nuclear Spectroscopic Telescope Array (NuSTAR) mission approximately two days after the glitch occurred. We also identified three other glitch events in archival observations from the CHIME/Pulsar instrument, which were fortuitously observed in X-rays by a different monitoring campaign with the Neutron star Interior Composition Explorer (NICER) mission. 

In this paper, we present an analysis of radio and X-ray emission from PSR~J2229+6114 surrounding four glitch events. In Section~\ref{sec:observation}, we describe observations obtained by two instruments at the CHIME telescope (CHAMPSS and CHIME/Pulsar) and two X-ray space missions (NuSTAR and NICER), along with the preprocessing procedures applied to prepare the data for further analysis. We then detail the detection and modeling of the four glitch events in Section \ref{sec:timing-analysis}. We show results from radio and X-ray emission analysis in Sections \ref{sec:results-radio} and \ref{sec:results-xray}, respectively, including X-ray flux analyses, pulse profile morphology in both bands, and searches for burst activity. In Section \ref{sec:discussion}, we present our result alongside four previous post-glitch X-ray follow-ups of RPPs and discuss the implications of these measurements for the unified neutron star model. We conclude and outline our future prospects in Section \ref{sec:conclusion-future-prospect}. 

\section{Observations and Preprocessing} \label{sec:observation}

\subsection{CHAMPSS and CHIME/Pulsar Observations}

CHIME is a transit telescope operating at a radio frequency of 400--800\,MHz \citep{chime_overview}. Our radio observations are recorded by the CHAMPSS and CHIME/Pulsar, two independent digital instruments at the telescope. 

CHAMPSS is a pulsar survey instrument that uses the beamformed intensity data stream from CHIME/FRB at a time resolution of 0.983\,ms. It stitches together the time series recorded from adjacent CHIME/FRB steady beams along the east-west direction to perform a pulsar periodicity search. Starting with the stitched time series, we can follow up known transiting sources by dedispersing the data incoherently at a known dispersion measure (DM) and folding the resulting data product into an `archive' file \citep[see][for technical details]{champss_overview}. For PSR~J2229+6114, the instrument records data for $\sim$13.4 minutes during the transit of the source in the CHIME field of view. The folded data archive for the source contains a 3D cube of 1,024 frequency channels, 81 time subintegrations, and 32 phase bins in PSRCHIVE format \citep{psrchive}. CHAMPSS recorded data for the source from 2026 January 21 to 2026 April 9 (MJDs 61061 to 61139). 

CHIME/Pulsar is a dedicated pulsar observing instrument that forms digital tracking beams from CHIME raw voltage (baseband) data. After beamforming, it coherently dedisperses the beamformed baseband data at a known DM and folds the data into PSRCHIVE archive files with a fixed resolution of 1,024 frequency channels \citep[see ][for technical details]{chimepsr_overview}. The instrument records data for PSR~J2229+6114 during the source's full transit in the CHIME field of view for $\sim$30~minutes (i.e., the time that the source drifts across the full-width at half-maximum of the CHIME beam at 400\,MHz). It has then been configured to save `archive' files with 256 rotational phase bins, 59 subintegrations, and four polarization products of XX$^*$, YY$^*$, Re(XY$^*$), Im(XY$^*$) for the source. \hl{CHIME/Pulsar has started observing the source on 2020 January 7 (MJD 58855), and we used all the data through 2024 July 22 (MJD 60513).
}


Folded archives from both digital instruments are further reduced by a pipeline processing to extract times of arrival (TOAs) for subsequent timing analysis. Radio-frequency interference (RFI) in the data is first mitigated using the \texttt{clfd} package \citep{clfd}. Then, the data cube in each archive is summed over polarization (for CHIME/Pulsar data), frequency, and time to produce a 1D pulse profile using the \texttt{pam} program in PSRCHIVE. We obtained a standard profile template by fitting von Mises profiles on an observed profile using the \texttt{paas} program. We subsequently used the Fourier domain with the Markov Chain Monte Carlo (FDM) shift algorithm embedded in the PSRCHIVE \texttt{pat} program to extract TOAs. Each TOA in our timing dataset corresponds to a single observation recorded during the telescope's daily transit.

\subsection{NuSTAR Observations} \label{sec:nustar_obs}

NuSTAR observed PSR~J2229$+$6114 through a triggered ToO observation starting on 2026 March 11 at 21:00:10 UTC (MJD 61110.88; ObsID 81102350002), roughly two days after the detection of the glitch. To compare the post-glitch emission with the pre-glitch levels, we also downloaded the only archival NuSTAR observation available on the HEASARC Browse interface\footnote{\url{https://heasarc.gsfc.nasa.gov}. } from 2020 September 23 (MJD 59113.99; ObsID 40660001002). 

The standard pipeline tools (\texttt{nupipeline} and \texttt{nuproducts}) in NuSTARDAS v2.1.5 were used for data reduction, using the CALDB 20260223 calibration. The final source and background event files were generated with circular regions with $30\arcsec$ radius around the source (22h 29m 0.5260s, $61\degr\, 14\arcmin\, 08.480\arcsec$) and $80\arcsec$ in a source-free region near the source, respectively. The total filtered exposure times for the archival (2020 September 23) and recent (2026 March 11) observations are approximately 45\,ks and 25\,ks, respectively. 


\subsection{NICER Observations} \label{sec:nicer_obs}

We used public archival data\footnote{Observations were obtained from the General Observer (PI: W. Ho) and Director's Discretionary Time programs.} from the NICER mission for our X-ray analysis, available on the HEASARC Browse interface. The available X-ray observing epochs range from MJDs 58872 to 60097 (ObsIDs 2579050915 to 6033370220), which partially overlap with our CHIME/Pulsar timing observations. 

Those observations were processed using NICER's standard processing pipelines from HEASoft\footnote{Version 6.35.1; \url{https://heasarc.gsfc.nasa.gov/docs/software/lheasoft/}. } and NICERSoft\footnote{Accessed on March 5, 2025; \url{https://github.com/paulray/NICERsoft}. }. We first applied standard screening and calibration procedures to produce cleaned event files using \texttt{nicerl2} \footnote{Version 2025-03-11\_V013a; \url{https://heasarc.gsfc.nasa.gov/docs/nicer/analysis\_threads/nicerl2/}. }. Then, we used \texttt{psrpipe.py} to filter the cleaned event files further with the following criteria: (a) overshoot count rate below 1.5 c/s per detector, (b) undershoot count rate below 600 c/s per detector, (c) space weather index below 5, and (d) magnetic cutoff rigidity of Earth's magnetic field above 1.5 GeV/c. 


\section{Timing Analysis} \label{sec:timing-analysis}

\subsection{Glitch Detections}

We identified all four glitch events in PSR~J2229+6114 through our glitch monitoring campaign. To enable near-real-time detection of glitches for the campaign, we built a glitch alert system based on the capability of the CHAMPSS Timing Pipeline to perform algorithmic pulsar timing \citep[see Section 4.3 of ][for technical details]{champss_overview}. The pipeline automatically processes CHIME observations and extracts pulse TOAs. Then, it iteratively fits timing models to TOAs and adds parameters based on the F-test \citep{apt}. In this pipeline process, glitches are identified as sudden losses of phase coherence in timing solutions from one day to the next, triggering real-time alerts.

We identified the first three glitches in the CHIME/Pulsar dataset during the commissioning testing of our glitch alert system. The first two glitches both disrupted phase coherence in the timing solution and were therefore detected immediately during the testing. The third glitch event was identified by an unmodeled early arrival of TOAs (i.e., a sudden spin-up in frequency) during modeling of the relaxation behavior of the second glitch. The long-term timing solution reveals that the frequency change induced by the spin-up is enduring, thereby distinguishing the event from red-noise effects. 

After the glitch alert system was commissioned, the fourth glitch occurred, manifesting as a sudden loss of phase coherence in the first post-glitch TOA extracted from the CHAMPSS observation. Our alert system detected the phase discontinuity and issued an alert $\sim$30\,minutes after the source's first transit following the glitch epoch at the CHIME telescope. The glitch event was confirmed by data recorded during subsequent transits after the glitch epoch. 

\subsection{Glitch Sizes} \label{sec:glitch-sizes}

We measured glitch sizes by obtaining local timing solutions from TOAs within $\pm 1$ month of the four glitch epochs and fitting simple glitch models that include only the glitch epoch, the change in frequency ($\Delta\nu$), and the change in spin-down rate ($\Delta\dot\nu$), without any recovery components. 

The post-fit parameters are presented in Table~\ref{tab:glitch-parameters}. The pre-glitch spin-down parameters ($\nu$ and $\dot\nu$) were fitted using PINT \citep{pint} with an initial timing solution from the ATNF Pulsar Catalogue\footnote{Version 2.6.5; \url{https://www.atnf.csiro.au/research/pulsar/psrcat/}. } \citep{psrcat}, using JPL DE405 Planetary Ephemerides and TT(TAI) time standard as implemented in PINT. Starting from pre-glitch models, the glitch parameters are derived using the Markov Chain Monte Carlo (MCMC) fitting routine embedded in PINT with \texttt{emcee} \citep{emcee} to properly sample degeneracies in glitch parameters. The reported glitch parameters in Table~\ref{tab:glitch-parameters} are therefore the medians, with uncertainties corresponding to the 16th–84th percentile interval, of the posterior distribution. The glitch epoch was included as a free parameter in the fit, with a uniform prior to constrain it to lie between the two consecutive TOAs before and after the glitch. The phase change at the glitch epoch is enforced to be zero. 

There were ambiguities in our measurement of spin frequencies and glitch sizes. Because CHIME's transit nature yields TOAs sampled uniformly at one per sidereal day, $\nu$ values that differ by $\pm N/T_{\rm sid}$ are not distinguishable using phase-coherent timing, where $N$ is an integer and $T_{\rm sid}$ is the sidereal day. The same ambiguity applies when measuring glitch amplitudes, as $\Delta\nu$ may be comparable to or larger than $1/T_{\rm sid}$. We applied the technique described by \cite{champss_overview} to all reported pre- and post-glitch timing solutions to ensure that the measured $\nu$ and $\Delta\nu$ values are not aliased. This technique first aligns multiple days of fold-mode observations using phase-coherent timing solutions. Then it stacks these observations in time--phase space to achieve a higher signal-to-noise ratio and measures the drift of pulse phases across individual transits induced by the aliased solutions. 

Among the four glitch events reported in the table, the first glitch (at MJD 59267) was measured previously by \cite{ggy+22}, while the remaining three glitches are presented here for the first time. The size of the first glitch is similar to that reported previously, despite a small discrepancy that is primarily due to the two independent measurements fitted with a simple glitch model without exponential recovery to TOAs spanning different time intervals. The simple model effectively approximates the underlying exponential with a linear model and therefore measures $\Delta\nu$ and $\Delta\dot\nu$ slightly differently depending on which portion of the post-glitch behavior is sampled. 

\begin{figure*}
    \centering
    \includegraphics[width=\linewidth]{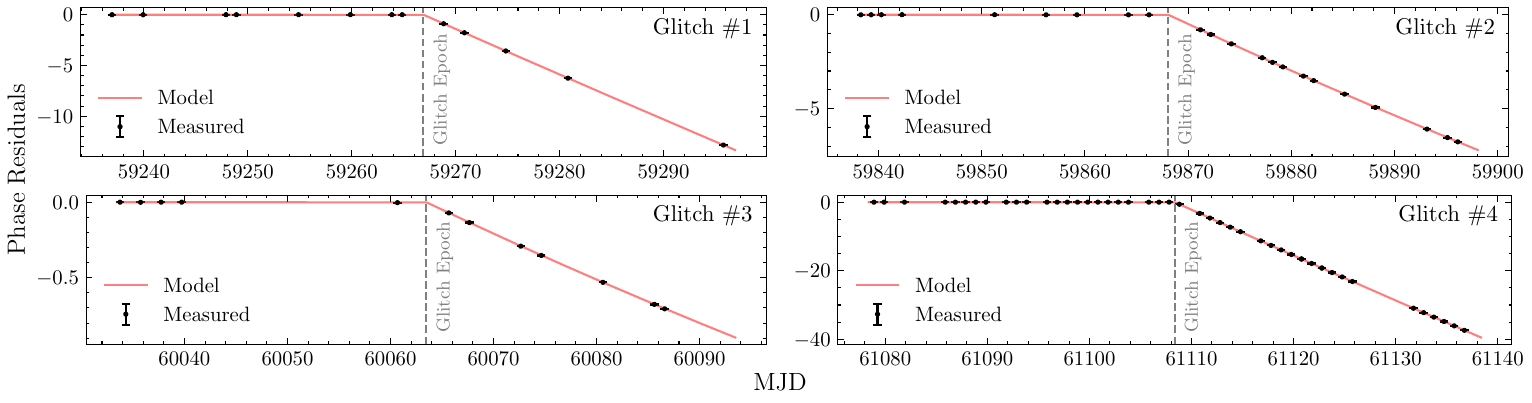}
    \caption{Timing residuals surrounding glitch epochs for PSR$~$J2229$+$6114. Each panel corresponds to a glitch identified in our timing observations with the CHIME telescope, with glitch epochs indicated by dashed vertical lines. Simple glitch models that include only glitch epoch, $\Delta\nu$, and $\Delta\dot\nu$ are overplotted as a red solid line in the pre-fit residuals. }
    \label{fig:timing-residuals}
\end{figure*}

\input{tb_glitch_parameters_simple}

Figure~\ref{fig:timing-residuals} shows residuals of TOAs surrounding all four glitch events compared to the pre-glitch models (i.e., the models without glitch parameters), overplotted with the fitted simple glitch models. All four glitches exhibit a sudden increase in the spin frequency (spin-up). The sizes of glitches 1, 2, and 4 are an order of magnitude larger than that of the third event, resulting in a sudden loss of phase coherence in the timing solution at the time of detection (this phase wrap is not visible in the figure as we manually aligned the TOAs for visualization). The first three glitches exhibit notable long-term recovery in our archival CHIME/Pulsar data, which we discuss further in Section~\ref{sec:glitch-rec}. 

\subsection{Glitch Recoveries} \label{sec:glitch-rec}

\input{tb_glitch_parameters_rec}

\begin{figure*}
    \centering
    \includegraphics[width=\linewidth]{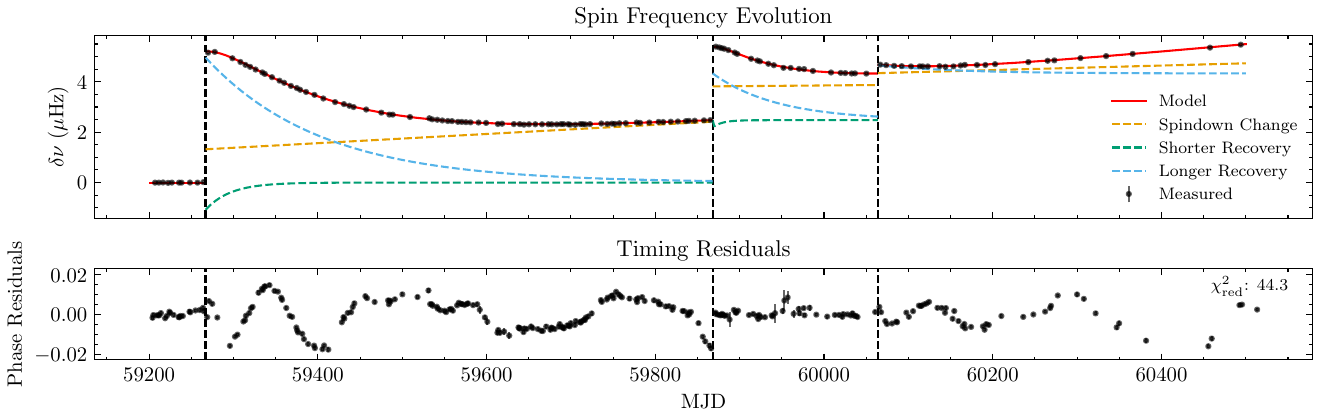}
    \caption{Spin frequency evolution of PSR$~$J2229$+$6114 and its timing residuals in phase compared to glitch models with exponential recovery components. In the top panel, the frequency evolution is presented as the change in spin frequency relative to the pre-fit model prior to glitch 1.  The measured spin frequencies are shown as black dots, and the post-fit glitch model is shown as a red solid line underneath the black dots. Contributions from individual model components are indicated in colored dashed lines, as specified in the legend. Since the model for glitch 3 has only one long-term recovery component, no green dashed line (i.e., the shorter recovery) is plotted for the glitch. The post-fit timing phase residuals are shown in the bottom panel. The post-fit reduced $\chi^2$ is indicated in the top-right corner of the panel. Glitch epochs are shown by dashed black lines in both panels. }
    \label{fig:timing-residuals-rec}
\end{figure*}

The CHIME/Pulsar dataset for glitches 1--3 spans from several months before the first glitch epoch and roughly a year after the third, and those three glitches exhibit evidence of long-term recovery in the dataset. We subsequently modeled each glitch's recovery using the full set of available CHIME/Pulsar TOAs with the longer timing baseline. Since the fourth glitch is newly discovered, it is not possible to perform a similar modeling due to its short post-glitch timing baseline at the time of writing. 

Table~\ref{tab:glitch-parameters-rec} shows post-fit glitch models with exponential decay terms ($\tau_{\rm d}$ and $\nu_{\rm d}$) for glitch recoveries. Using those simple glitch models in Section~\ref{sec:glitch-sizes} as a starting point, those models are fitted jointly by phase-connecting all the TOAs from MJD~59204 to 60513. Model parameters are again derived with MCMC, and the reported values are posterior medians with 16th–-84th percentile uncertainties. Glitches 1 and 2 exhibit more complex recoveries, so we fitted them with two sets of decay terms on different timescales. The relaxation of glitch 3 is simpler and can be well described by a single decay term. However, since the second glitch was interrupted by the third, parameter degeneracies leave the second glitch's longer recovery timescale poorly constrained. Hence, we fixed the timescale at 76.5 days (varying it by $\pm$1 day did not significantly affect the remaining parameter estimates).

To visualize those glitch models, we derived spin-frequency evolution from the phase offsets between adjacent TOAs (rather than measuring it directly from individual observations due to insufficient precision). The derived spin frequency evolution, along with the evolution predicted by glitch recovery models, is shown in the top panel of Figure~\ref{fig:timing-residuals-rec}. The first two glitches exhibit a short-term spin-up recovery soon after the glitch epoch, but later the recovery is dominated by the long-term spin-down recovery. The third glitch also shows a relaxation soon after the sudden spin-up during the relaxation of the second glitch. The bottom panel of the figure shows the post-fit timing residuals of the glitch models. The fit has a reduced $\chi^2$ of 44.3, with the excess being primarily due to variations in the residuals that are low amplitude, likely due to inexact glitch modeling, and a smaller contribution from red noise. Nevertheless, those variations are centered at zero with an rms scatter of 0.006 in phase (0.6\%), indicating a valid fit.

\section{Radio Emission} \label{sec:results-radio}

\subsection{Radio Profile Morphology}

\begin{figure}
    \centering
    \includegraphics[width=1\linewidth]{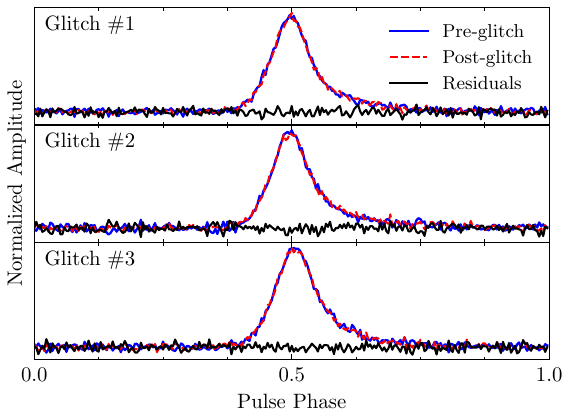}
    \caption{Radio pulse profiles of PSR$~$J2229$+$6114 averaged over 5 days of CHIME/Pulsar observations before and after the respective glitch, where each panel corresponds to profiles around a different glitch. In each panel, the pre-glitch and post-glitch average profiles are shown as blue solid and red dashed lines, respectively. The black solid line represents the residual between the two profiles, defined as the difference in amplitude between the post- and pre-glitch profiles. }
    \label{fig:radio-pp}
\end{figure}


Figure~\ref{fig:radio-pp} shows 5 day averaged pulse profiles before and after glitches 1--3 from 400--800\,MHz, where pre-glitch, post-glitch, and residuals are shown in each panel that corresponds to a detected glitch. To first obtain daily pulse profiles, we updated our fold-mode data using local timing solutions fitted by TOAs near glitch epochs, then averaged the frequency and time information using PSRCHIVE's \texttt{pam} routine. We subsequently used the \texttt{psradd} routine to sum the aligned daily profiles. \hl{
We subtracted the baseline on each summed profile using the built-in function in PSRCHIVE. Since our profile amplitudes are not flux-calibrated, we rescaled each post-glitch profile by the factor that minimizes the squared residual relative to the corresponding pre-glitch profile (i.e., by a factor of $\sum_i p_{0, i} p_{1, i}/\sum_i p^2_{1, i}$, where $p_{0, i}$ and $p_{1, i}$ are the pre-glitch and post-glitch profile amplitudes at bin $i$). The residuals shown in the figure are calculated by subtracting the pre-glitch profile from the post-glitch profiles. The reduced $\chi^2$ values for the on-pulse region in residuals (with a null hypothesis that the residuals are consistent with white Gaussian noise) in all three glitches are $\sim$1, with a number of degrees of freedom of 100 (number of on-pulse bins minus 2), indicating no statistically significant variation in radio profiles before and after the glitches. }

\begin{figure*}
    \centering
    \includegraphics[width=1\linewidth]{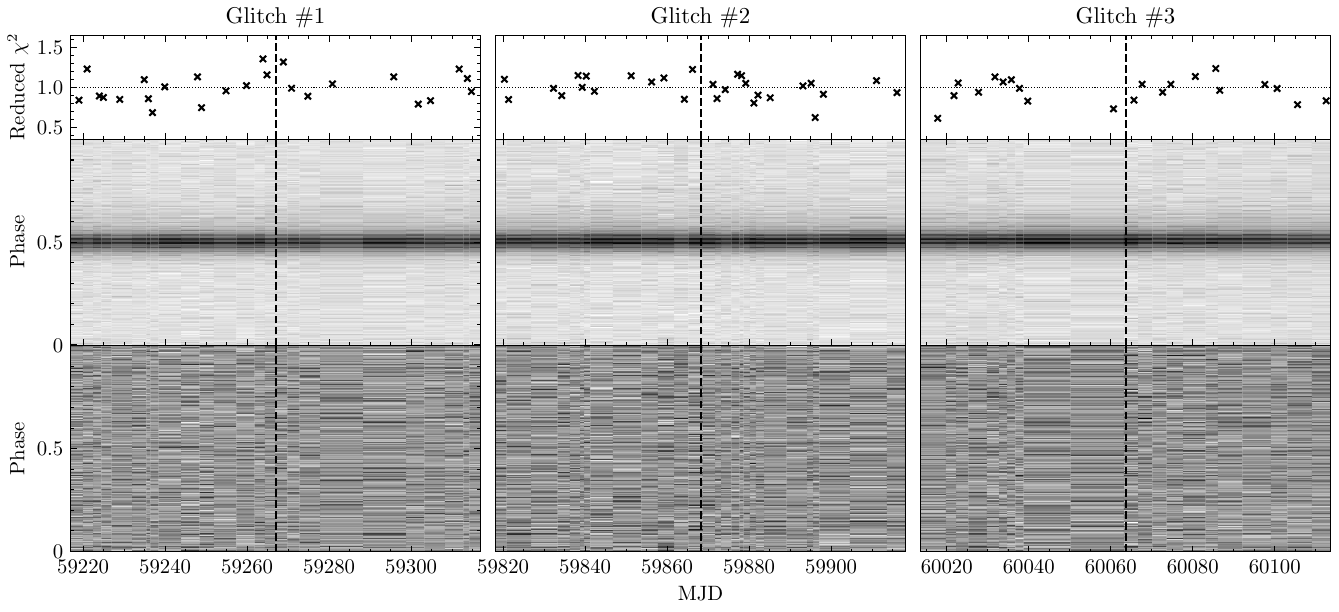}
    \caption{\hl{Radio pulse profile variations of PSR$~$J2229$+$6114 in CHIME/Pulsar observations surrounding the epochs of glitch 1--3, with each subfigure for a different glitch. The middle and bottom panels in each subfigure show profiles and profile residuals (relative to the pre-glitch profile morphology) over time. The top panel presents the reduced $\chi^2$ values of the on-pulse region in residuals as a function of time, with the horizontal dashed line denoting the reduced $\chi^2$ at one. With 100 degrees of freedom, a reduced $\chi^2$ of 1.25 deviates from the white-noise assumption at the $p < 0.05$ level, indicating statistically significant profile variation. The vertical dashed line indicates the epoch of each glitch.  }}
    \label{fig:radio-pp-var}
\end{figure*}

\hl{

The long-term averaged pulse profiles in RPPs are usually highly stable. To identify potential transient morphological variations in short timescales, we compared individual profiles with pre-glitch morphology and plotted them in Figure~\ref{fig:radio-pp-var}. We have again updated the fold-mode data surrounding three glitch epochs using the local timing solutions and averaged them per day to obtain daily pulse profiles. For each glitch, to create a template for pre-glitch profile morphology, we summed individual pre-glitch profiles in Figure~\ref{fig:radio-pp-var} with PSRCHIVE's \texttt{psradd}, then smoothed them with \texttt{pas}. Then, for each individual profile, we removed the baseline, scaled the template to match the profile's amplitude, as before, and computed the difference between the two profiles as the profile residuals. Finally, we again computed the reduced $\chi^2$ values for the on-pulse region of each residual. The reduced $\chi^2$ values in Figure~\ref{fig:radio-pp-var} vary similarly before and after three glitches. There are two observations surrounding glitch 1 that show reduced $\chi^2$ values that deviate from the noise by a statistically significant amount. However, since there is no clear structure in their corresponding residuals, we attribute the excessive $\chi^2$ values to stochastic profile variations (e.g., pulse jitters). 

}

A similar profile analysis is not possible for the fourth glitch because high-time-resolution timing observations of the source were not recorded prior to the glitch. Pre-glitch fold-mode observations for glitch 4 were recorded in 32 rotational phase bins by the CHAMPSS instrument, whereas CHIME/Pulsar observations with finer time resolution of 256 bins began only after the glitch was detected.
 
\subsection{Radio Bursts}

In addition to analyzing the fold-mode data from CHIME/Pulsar, we also searched the CHIME/FRB database for bright single radio pulses from the source. CHIME/FRB is an all-sky, real-time survey for dispersed single pulses using the CHIME telescope, and focuses primarily on fast radio bursts (FRBs) discovery \citep[see][for technical details]{chimefrb-overview}; it is also sensitive to giant radio pulses and single radio bursts from pulsars and magnetars (see \citealt{good+21} and \citealt{chime_magnetar}, for examples). By default, CHIME/FRB does not save raw data for known Galactic burst events, including pulsars, but it retains header information for every detected event in its database. The header information allows us to search for bursting behavior in PSR~J2229+6114 by identifying an excess of events coincident with the source's sky position and DM values surrounding the glitch epoch. 

We queried the CHIME/FRB database for single-pulse events registered between 2017 February 1 and 2026 April 1. Then we filtered these events using the following two criteria: (a) events labeled as PSR~J2229$+$6114 by CHIME/FRB's Known Source Sifter \citep[see Section 4.5 of][]{chimefrb-overview}, and (b) events within a positional offset of $\pm 5$\degr\ and a DM offset of $\pm 5\,{\rm pc\ cm^{-3}}$ from the pulsar. Our database query returned 172 candidate events, including some detected near the epoch of glitch 1. However, we determined that these events are spurious because their detection rates correlate with those at nearby sky positions, indicating poor localization, typical of RFI. Without any raw data being recorded, no further analysis can be done to verify the nature of these candidates. Thus, we conclude that our database query revealed no radio bursts clearly associated with the pulsar. 

\section{X-ray Emission} \label{sec:results-xray}

\subsection{X-ray Fluxes}

\begin{figure*}
    \centering
    \includegraphics[width=\linewidth]{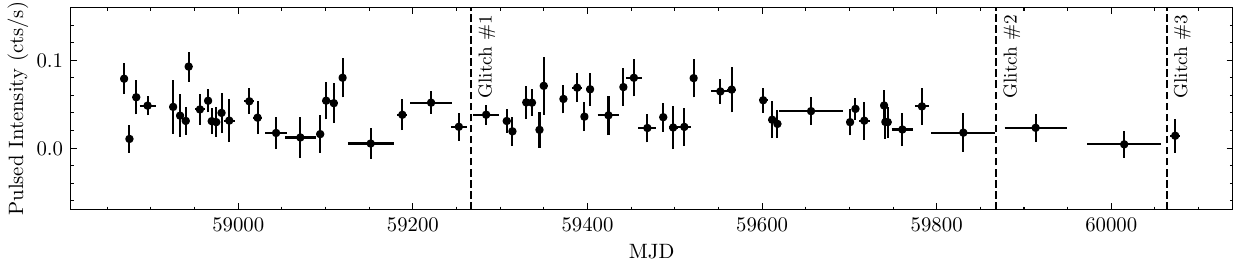}
    \caption{Pulsed 0.5--10-keV X-ray flux evolution of PSR~J2229$+$6114 from archival NICER observations, covering glitches 1--3. Epochs of glitch events are indicated by vertical dashed lines. Multiple observations are combined into a single measurement to achieve at least 10\,ks of exposure per data point. The time spans of the combined observations are indicated by the horizontal error bar.  }
    \label{fig:xray-flux}
\end{figure*}


From the NICER dataset, we combined multiple observations into a single measurement to achieve a minimum exposure of 10\,ks per data point, and measured the pulsed X-ray fluxes of the source over a 3 yr time span covering glitches 1--3. Those measurements are plotted as a function of time in Figure~\ref{fig:xray-flux}. \hl{The reduced $\chi^2$ test indicates the pulsed flux is broadly consistent with the weighted mean of $0.041 \pm 0.002$\,cts/s ($\chi^2_{\rm red}=1.24$, $p=0.09$).} The 3$\,\sigma$ 0.5--10\,keV upper limits for X-ray emission enhancement in glitches 1, 2, and 3 are 0.06 cts/s, 0.08 cts/s, and 0.07 cts/s, respectively (corresponding to a fractional enhancement up to 2.4, 4.6, and 17.3 above the pre-glitch level, respectively, calculated from the nearest data point before and after the glitch in Figure~\ref{fig:xray-flux}). \hl{These upper limits show no statistically significant flux enhancement at the glitch epochs.} If PSR~J2229+6114 showed an enhancement in X-ray emission after glitches 1--3 at a similar level to the two previously detected radiative changes in a high-$B$ pulsar, our dataset would have been sensitive enough to detect it (e.g., PSR~J1846$-$0258 showed a fractional X-ray flux change $\Delta F_{\rm X} / F_{\rm X} \sim 7.7$).  However, it should be noted the bin sizes in Figure~\ref{fig:xray-flux} for the first data point after glitches 1, 2, and 3 are 15.2 days, 35.5 days, and 5.55 days, respectively; if a hypothetical flux enhancement had decayed on a very short timescale that is much shorter than bin sizes, we may not be able to capture the emission change in this NICER dataset. 


\input{tb_nustar_flux}

\begin{figure}
    \centering
    \includegraphics[width=\linewidth]{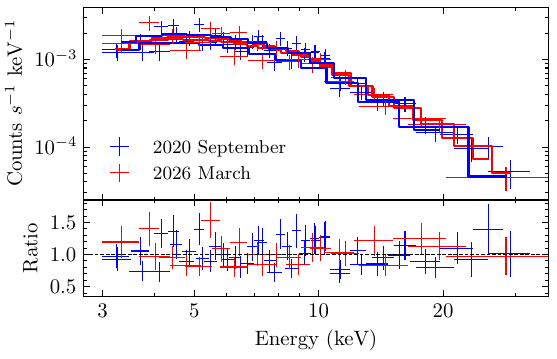}
    \caption{\hl{X-ray spectra of PSR~J2229+6114 from two epochs of NuSTAR observations. The top panel presents the spectra from FPMA and FPMB before (blue) and after (red) glitch 4. The data for both epochs were fitted with a \texttt{tbabs*pow} model (see Table~\ref{tab:nustar_flux}); the best-fit models are shown as solid lines. The bottom panel shows the ratio (unitless) of data to model (data/model), with a horizontal dashed line indicating a ratio of one. }}
    \label{fig:nustar-spec}
\end{figure}


For the two NuSTAR observations, we extracted the spectra with \texttt{nuproducts} (see Section~\ref{sec:nustar_obs}) and fitted them with an absorbed power-law model using XSPEC v12.15.1 \citep{arnaud96}. \hl{The fitted spectral parameters are listed in Table~\ref{tab:nustar_flux} and the spectra are plotted in Figure~\ref{fig:nustar-spec}.} We used the 2020 September observation as our reference spectrum for the era before glitch 4, as the observation (MJD 59113) was not known to be close in time to any known glitch in PSR~J2229$+$6114 \citep{bsa+22}.  The best-fit power-law indices ($\Gamma_{\rm X}$) are consistent within $1\,\sigma$, indicating no statistically significant spectral evolution between the two epochs. The post-glitch 3--79\,keV absorbed flux from the source is marginally lower than the pre-glitch level, with a difference of $\Delta F_{\rm X}^{\rm abs}=-1.0(3)\times 10^{-12}$\,erg\,s$^{-1}$\,cm$^{-2}$. We therefore conclude that there is no evidence for an X-ray flux enhancement following glitch 4 in NuSTAR observations. To allow for comparison with previous measurements from other sources in Section~\ref{sec:discussion}, we place a 3$\,\sigma$ upper limit on the unabsorbed flux enhancement in the 0.5--10\,keV band of $0.03 \times 10^{-12}$\,erg\,s$^{-1}$\,cm$^{-2}$ (corresponding to a fractional enhancement up to 0.03).

\subsection{X-ray Bursts}

Some magnetars and high-$B$ RPPs show short X-ray bursts \citep[e.g.,][]{kgw+03,J1846} after glitch events. For this reason, we extracted the X-ray light curve with 5-ms time bins from the NICER (0.1--10\,keV) and NuSTAR (3--79\,keV) datasets and searched for short X-ray bursts exceeding the mean count rate by $> 5\,\sigma$, assuming Poisson statistics. Our search identified seven burst-like candidate events in NICER observations, but they are likely not astrophysical. All seven candidates were detected by only one detector (of the 52 active detectors on board NICER), indicating they are likely due to instrumental artifacts or particle background. The NuSTAR search returned no candidates above threshold. We therefore find no significant detections of X-ray bursts that are clearly associated with the source.

\subsection{X-ray Profile Morphology} \label{sec:x-ray-profiles}

\begin{figure}
    \centering
    \includegraphics[width=\linewidth]{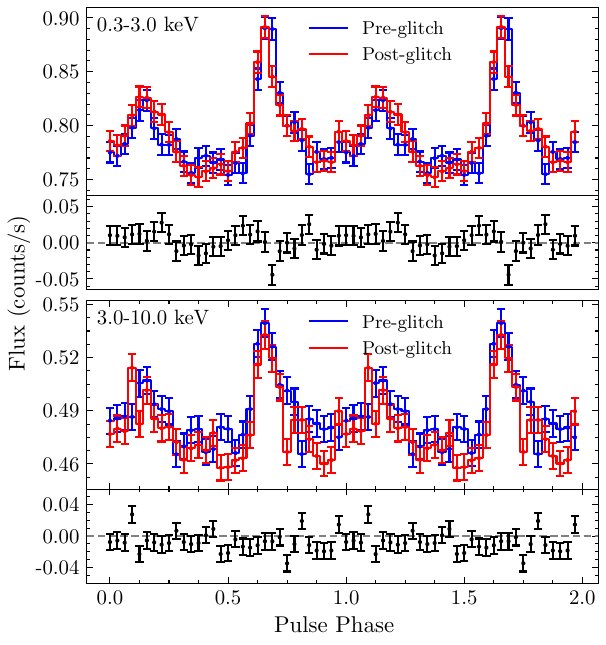}
    \caption{X-ray pulse profiles of PSR$~$J2229$+$6114 accumulated over $\sim$1 year of NICER observations before and after glitch 1. Profiles and their residuals in soft (0.3--3.0\,keV) and hard (3.0--10.0\,keV) \hl{X-rays} are plotted in the top and bottom subfigures, respectively. In each subfigure, the top panel shows the pre-glitch (blue) and post-glitch (red) profiles. The bottom panel shows the absolute residuals, obtained by subtracting the pre-glitch profile from the post-glitch profile without normalization or median subtraction. Each profile is repeated twice for better visualization. }
    \label{fig:xray-profile}
\end{figure}

\begin{figure}
    \centering
    \includegraphics[width=\linewidth]{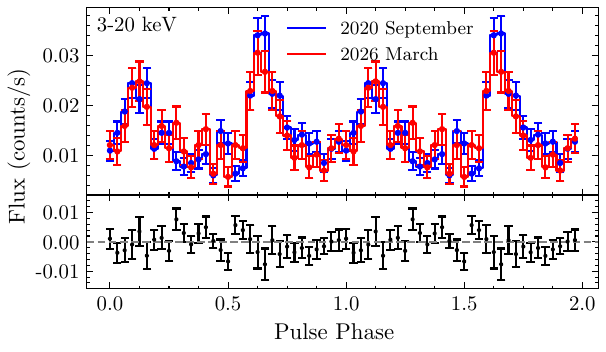}
    \caption{\hl{X-ray pulse profiles of PSR~J2229$+$6114 from two epochs of NuSTAR observations. The top panel presents the profiles before (blue) and after (red) glitch 4 in hard X-rays (3--20$\,$keV). The bottom panel shows the absolute profile residual. Each profile is repeated twice for better visualization. }}
    \label{fig:xray-profile-nustar}
\end{figure}

\hl{We folded NICER and NuSTAR observations to produce X-ray pulse profiles for PSR~J2229+6114 before and after glitches 1 and 4, presented in Figures~\ref{fig:xray-profile} and \ref{fig:xray-profile-nustar}, respectively. A similar analysis is not possible for the remaining glitches 2 and 3 because they are too close in time; the surrounding NICER observations provide insufficient exposure to construct clear profiles. }

\hl{
In order to produce the 0.3--10\,keV X-ray profiles presented in Figure~\ref{fig:xray-profile}, we averaged over $\sim$1 yr of NICER observations before and after glitch 1. To phase-align NICER observations, we obtained a separate timing solution using radio TOAs from CHIME/Pulsar, covering the NICER observing epochs. This solution was derived by starting with our glitch models in Table~\ref{tab:glitch-parameters} and fitting higher-order spin-frequency derivatives to suppress red noise in the timing residuals. While this approach introduces timing artifacts, these do not affect our X-ray analysis, which relies on phase alignment rather than timing precision.  We subsequently calculated each photon's rotational phase with the ephemeris using \texttt{photonphase} in PINT \citep{pint} and binned the photons into pulse profiles.

} 

We subtracted the two \hl{NICER pulse profiles} and computed the reduced $\chi^2$ of the resulting differences relative to zero (i.e., white Gaussian noise). In the 0.3--3.0\,keV band, the residuals are consistent with Gaussian noise, indicating no statistically significant variation ($\chi^2_{\rm red}=1.12$, $p=0.268$) \hl{for glitch 1}. Notably, despite the source exhibiting visually similar hard X-ray profiles before and after the glitch, residual analysis in the 3.0--10.0\,keV band reveals a modest excess variation inconsistent with white noise ($\chi^2_{\rm red}=1.94$, $p=0.001$), suggesting possible evolution in the hard X-ray emission. However, it is possible that this excess variation is primarily due to the pre-glitch baseline flux being slightly brighter. If we normalize the two profiles in the hard X-ray band by subtracting the median fluxes from each phase bin, the residuals analysis again shows no statistically significant variation in their morphology ($\chi^2_{\rm red}=1.42$, $p=0.057$) \hl{across glitch 1}. With this dataset, however, we cannot determine whether the baseline offset arises from the source or the background emission.

\hl{
To produce the 3--20\,keV X-ray profiles from NuSTAR, we again calculated the rotational phase of each photon using \texttt{photonphase}. For the pre-glitch epoch, we used a local timing solution fitted from CHIME/Pulsar radio TOAs surrounding the NuSTAR observation, starting from the pre-glitch timing solution of glitch 1 in Table~\ref{tab:glitch-parameters}. For the post-glitch epoch, we used the local timing solution for glitch 4 in Table~\ref{tab:glitch-parameters}. In both models, we fitted the phase offset (relative to the radio pulse profile template from which TOAs are extracted) to align profiles from two epochs at the same rotational phase. We generated the pulse profile and residuals as before. The reduced $\chi^2$ of the residuals suggests no statistically significant variation in profiles between the two epochs of NuSTAR observations ($\chi^2_{\rm red}=0.99$, $p=0.49$). 
}

\section{Discussion} \label{sec:discussion}

\input{tb_det_summary}

\hl{Our analysis revealed no evidence of X-ray flux enhancement or short X-ray bursts following any of the four glitches in PSR~J2229+6114. These results increase the number of rapid post-glitch X-ray observations of RPPs from four to five.} In this section, we present our results alongside previous similar measurements and discuss their implications in the context of the unifying neutron star model, in which the $B$-field is the unifying parameter. 

Table~\ref{tab:det_summary} summarizes all rapid X-ray follow-ups of glitches in five RPPs to date. The two high-$B$ RPPs exhibited magnetar-like outbursts immediately after glitch events, suggesting that these high-$B$ RPPs are transitional objects between RPPs and magnetars: PSR$~$J1846$-$0258 ($B_{\rm dip} \approx  5 \times 10^{13}$\,G; \citealt{gvb+00}) exhibited two large glitches in 2006 May and 2020 August, each were accompanied by a magnetar-like outburst that lasted for $\sim 2$ months \citep{J1846, J1846_2020}. PSR$~$J1119$-$6127 ($B_{\rm dip} \approx 4 \times 10^{13}$\,G; \citealt{ckl+00}) showed a similar outburst in 2016 July immediately after a large glitch with $\Delta \nu / \nu = 5.74(8) \times 10^{-6}$ \citep{J1119}. Both sources also exhibited a temporary transition to a magnetar-like X-ray spectrum during their outbursts \citep[see discussion by][for a review]{J1119}. Indeed, the inferred $B$-field strength could serve as a unifying parameter for neutron stars, thereby bridging or perhaps unifying the perceived dichotomy between RPPs and magnetars in neutron star astrophysics \citep[see][for a review]{kas03}. 

Despite the detection of X-ray burst activity in two high-$B$ RPPs at glitch epochs, confirmation of a unified picture requires consideration of behavior near the lower end of the high-$B$ regime. It remains unclear whether this magnetar-like behavior is a general characteristic of most of the population and whether it scales with the inferred surface magnetic field strength. However, rapid post-glitch X-ray follow-up of lower-$B$-field RPPs ($B_{\rm dip}<10^{13}$\,G) has been rare due to observational challenges, including the need for high-cadence monitoring, despite more pulsars being discovered in the transitional regime than in the high-$B$ regime. Prior to the NuSTAR follow-up of PSR~J2229$+$6114 reported by this paper, such observations have thus far only occurred in the Vela pulsar ($B_{\rm dip} \approx 3 \times 10^{12}$\,G; \citealt{vela}) and the Crab pulsar ($B_{\rm dip} \approx 4 \times 10^{12}$\,G; \citealt{sls+18, crab})\footnote{Glitches in the Crab pulsar are special, exhibiting distinct properties such as delayed spin-ups, as well as correlated magnetospheric changes that suggest a possible magnetospheric contribution \citep[see][for a review]{zgy+22}.}, as listed in Table~\ref{tab:det_summary}. All follow-up observations were triggered by glitch detections from dedicated monitoring campaigns. A measurable glitch-induced change in X-ray emission was found in neither RPP, supporting the idea that glitch-related magnetar-like emission is likely unique to, or at least most common in, high-$B$ RPPs. 

It is worth noting that all three glitches from lower $B$-field RPPs in Table~\ref{tab:det_summary} have smaller fractional frequency changes ($\Delta\nu/\nu$) than the two glitches in high-$B$ RPPs, but their changes in rotational kinetic energy ($\Delta E_{\rm rot}$) are all greater than those seen in the two high-$B$ RPPs. This suggests that the bursting behavior in high-$B$ RPPs may be independent of the glitch’s change in rotational kinetic energy, despite their strong connection with glitch events. The lack of dependence on changes in rotational kinetic energy is consistent with the model prediction in \cite{pp+11} that the bursting behavior in neutron stars is instead driven by the elastic energy accumulated in their crust, which is probably released via starquakes. However, the mechanism linking bursting behavior to glitches remains an open question.  

The data currently support more dramatic post-glitch emission activity in high-$B$ RPPs, as in magnetars, but the statistics are still limited to constrain the underlying relationship between $B$-field and outburst. The X-ray follow-ups listed in Table~\ref{tab:det_summary} thus far are for sources with both conventional spin-inferred $B$-fields and for those with much higher values; how the post-glitch emission behavior in RPPs changes with $B$-fields in the intermediate -- likely transitional -- regime (from $4 \times 10^{12}$ to $4 \times 10^{13}$\,G) remains unclear. It is also unknown whether there are any hidden parameters, in addition to the $B$-field strength, that relate to the post-glitch emission activity in RPPs (e.g., multipolar fields, characteristic ages). Thus, future nondetections or weak detections of post-glitch emission in RPPs are essential for mapping any continuous relationship that could break the dichotomous classification of RPPs and magnetars in neutron star astrophysics.

\section{Conclusions and Future Prospects} \label{sec:conclusion-future-prospect}

We have presented the first result from our pulsar glitch monitoring campaign with the CHIME telescope, which analyzed the radio and X-ray emission surrounding four glitch events in PSR~J2229+6114. Among the four events, the first three events were identified in archival CHIME/Pulsar data, and the fourth event was detected by our glitch alert system immediately after its first transit since the glitch epoch. \hl{
Our analysis showed no evidence of X-ray flux enhancement following any of the four glitches in PSR~J2229+6114. We also found no changes in profile morphology (glitches 1 and 4 in X-rays and glitches 1–3 in the radio band) and no bursts in either band. These post-glitch radiative behaviors are in stark contrast to the significant changes in X-ray and radio emission observed after comparable glitches in RPPs with much higher magnetic fields. 

}


Using radio datasets, we measured the glitch epochs and sizes for all four glitch events and modeled the long-term relaxation behavior for the first three glitches with longer timing baselines. We also compared the morphology of the radio profiles before and after the first three glitch events using high-resolution CHIME/Pulsar observations and found no significant profile variations. To search for evidence of glitch-induced radio bursts, we queried the CHIME/FRB database and found that none of the candidate events detected by the system were clearly associated with the source. From X-ray datasets, we measured the flux evolution for all four glitch events, \hl{found no evidence for X-ray flux enhancement, and reported $3\,\sigma$ upper limits on any such enhancement}. We also searched for short X-ray bursts in X-ray light curves and found none that were clearly associated with the source's emission. Additionally, we compared the source's X-ray pulse profile before and after \hl{glitches 1 and 4}, finding no significant variations in profile morphology, except for a slight baseline offset \hl{in glitch 1's hard X-ray profile}, the origin of which could not be determined. 

The absence of radiative change after glitches in PSR~J2229+6114, an RPP with a lower $B$-field than PSRs~J1846$-$0258 and J1119$-$6127, is consistent with predictions of a unifying neutron star model in which the $B$-field strength serves as the unifying parameter. This supports the idea that post-glitch radiative changes in RPPs are probably unique to or at least most common among those with a high $B$-field. However, our discussion demonstrated that X-ray follow-up of RPPs after glitches remains worthwhile, particularly in the mid- to high-$B$ regime. Our pulsar glitch monitoring campaign is currently observing 20 high-$B$ RPPs with $B_{\rm dip}>5\times 10^{12}$\,G at daily cadence, and this number will grow as ongoing all-sky pulsar surveys in the Northern Hemisphere \citep[e.g., the CHAMPSS Survey; ][]{champss_overview} discover more new pulsars in the high-$B$ regime. In the future, we expect the campaign to capture large glitch events within 48 hr of their occurrence and to trigger rapid X-ray follow-up through ToO observations. It may also enable near-simultaneous radio and X-ray follow-up during glitch epochs. We hope this effort will improve our statistics and thereby impose further constraints on post-glitch behavior of those RPPs in the likely transitional $B$-field regime from $4 \times 10^{12}$ to $4 \times 10^{13}$\,G.

\begin{acknowledgments}

\allacks
\end{acknowledgments}

\bibliography{refs}{}
\bibliographystyle{aasjournal}




\end{document}

%% file: auth.tex
\author[0009-0009-9343-4193]{Wenke Xia}
  \affiliation{Department of Physics, McGill University, 3600 rue University, Montr\'eal, QC H3A 2T8, Canada}
  \affiliation{Trottier Space Institute, McGill University, 3550 rue University, Montr\'eal, QC H3A 2A7, Canada}
\author[0000-0002-7164-9507]{Robert Main}
  \affiliation{Department of Physics, McGill University, 3600 rue University, Montr\'eal, QC H3A 2T8, Canada}
  \affiliation{Trottier Space Institute, McGill University, 3550 rue University, Montr\'eal, QC H3A 2A7, Canada}
\author[0000-0002-0940-6563]{Mason Ng}
  \affiliation{Department of Physics, McGill University, 3600 rue University, Montr\'eal, QC H3A 2T8, Canada}
  \affiliation{Trottier Space Institute, McGill University, 3550 rue University, Montr\'eal, QC H3A 2A7, Canada}
\author[0000-0001-9345-0307]{Victoria M.~Kaspi}
  \affiliation{Department of Physics, McGill University, 3600 rue University, Montr\'eal, QC H3A 2T8, Canada}
  \affiliation{Trottier Space Institute, McGill University, 3550 rue University, Montr\'eal, QC H3A 2A7, Canada}
\author[0000-0003-2317-1446]{Jason W.~T.~Hessels}
  \affiliation{Department of Physics, McGill University, 3600 rue University, Montr\'eal, QC H3A 2T8, Canada}
  \affiliation{Trottier Space Institute, McGill University, 3550 rue University, Montr\'eal, QC H3A 2A7, Canada}
  \affiliation{Anton Pannekoek Institute for Astronomy, University of Amsterdam, Science Park 904, 1098 XH Amsterdam, The Netherlands}
  \affiliation{ASTRON, Netherlands Institute for Radio Astronomy, Oude Hoogeveensedijk 4, 7991 PD Dwingeloo, The Netherlands}
\author[0009-0007-0757-9800]{Alyssa Cassity}
  \affiliation{Department of Physics and Astronomy, University of British Columbia, 6224 Agricultural Road, Vancouver, BC V6T 1Z1 Canada}
\author[0009-0004-5775-8821]{Abigail K.~Denney}
  \affiliation{David A. Dunlap Department of Astronomy and Astrophysics, 50 St. George Street, University of Toronto, ON M5S 3H4, Canada}
\author[0000-0001-8384-5049]{Emmanuel Fonseca}
  \affiliation{Department of Physics and Astronomy, West Virginia University, PO Box 6315, Morgantown, WV 26506, USA }
  \affiliation{Center for Gravitational Waves and Cosmology, West Virginia University, Chestnut Ridge Research Building, Morgantown, WV 26505, USA}
\author[0000-0003-1884-348X]{Deborah C.~Good}
  \affiliation{Department of Physics and Astronomy, University of Montana, 32 Campus Drive, Missoula, MT 59812, USA}
\author[0009-0002-0330-9188]{Ajay Kumar}
  \affiliation{National Centre for Radio Astrophysics, Post Bag 3, Ganeshkhind, Pune, 411007, India}
\author[0000-0003-4634-5453]{Lars K\"{u}nkel}
  \affiliation{Department of Physics, McGill University, 3600 rue University, Montr\'eal, QC H3A 2T8, Canada}
  \affiliation{Trottier Space Institute, McGill University, 3550 rue University, Montr\'eal, QC H3A 2A7, Canada}
\author[0000-0001-8845-1225]{Bradley W.~Meyers}
  \affiliation{Australian SKA Regional Centre (AusSRC), Curtin University, Bentley WA 6102, Australia}
  \affiliation{International Centre for Radio Astronomy Research (ICRAR), Curtin University, Bentley WA 6102, Australia}
\author[0000-0002-8912-0732]{Aaron B.~Pearlman}
  \altaffiliation{NASA Hubble Fellow.}
  \affiliation{MIT Kavli Institute for Astrophysics and Space Research, Massachusetts Institute of Technology, 77 Massachusetts Ave, Cambridge, MA 02139, USA}
  \affiliation{Department of Physics, Massachusetts Institute of Technology, 77 Massachusetts Ave, Cambridge, MA 02139, USA}
  \affiliation{Department of Physics, McGill University, 3600 rue University, Montr\'eal, QC H3A 2T8, Canada}
  \affiliation{Trottier Space Institute, McGill University, 3550 rue University, Montr\'eal, QC H3A 2A7, Canada}
\author[0000-0001-9784-8670]{Ingrid Stairs}
  \affiliation{Department of Physics and Astronomy, University of British Columbia, 6224 Agricultural Road, Vancouver, BC V6T 1Z1 Canada}
\newcommand{\allacks}{
M.N. is a Fonds de Recherche du Quebec - Nature et Technologies (FRQNT) postdoctoral fellow. 
V.M.K. holds the Lorne Trottier Chair in Astrophysics \& Cosmology, a Distinguished James McGill Professorship, and receives support from an NSERC Discovery grant (RGPIN 228738-13).
The AstroFlash research group at McGill University, University of Amsterdam, ASTRON, and JIVE is supported by: a Canada Excellence Research Chair in Transient Astrophysics (CERC-2022-00009); an Advanced Grant from the European Research Council (ERC) under the European Union's Horizon 2020 research and innovation programme (`EuroFlash'; Grant agreement No. 101098079); an NWO-Vici grant (`AstroFlash'; VI.C.192.045); an ERC Starting Grant (`EnviroFlash'; Grant agreement No. 101223057); and an NWO-Veni grant (VI.Veni.222.295).
A.C. holds a UBC 4-Year Fellowship. 
E.F. is supported by the National Science Foundation (NSF) under grant number AST-2407399.
D.C.G. is supported by NSF Astronomy and Astrophysics Grant (AAG) award \#2406919.
A.B.P. acknowledges support by NASA through the NASA Hubble Fellowship grant HST-HF2-51584.001-A awarded by the Space Telescope Science Institute, which is operated by the Association of Universities for Research in Astronomy, Inc., under NASA contract NAS5-26555. A.B.P. also acknowledges prior support from a Banting Fellowship, a McGill Space Institute~(MSI) Fellowship, and a Fonds de Recherche du Quebec -- Nature et Technologies~(FRQNT) Postdoctoral Fellowship.
Pulsar research at UBC is supported by an NSERC Discovery Grant and by the Canadian Institute for Advanced Research. 
}

%% file: tb_glitch_parameters_simple.tex
\begin{deluxetable*}{cccccccccc}

\tablecaption{Glitches Identified in CHIME/Pulsar Observations} 
\tabletypesize{\footnotesize}
\label{tab:glitch-parameters}

\tablehead{
\colhead{\#} & \colhead{Period Epoch} & \colhead{$\nu$} & \colhead{$\dot\nu$} & \colhead{Glitch Epoch} & \colhead{$\Delta \nu$} & \colhead{$\Delta \dot\nu$} & \colhead{$\Delta \nu/\nu$} & \colhead{$\Delta \dot\nu/\dot\nu$} & \colhead{$\Delta E_{\rm rot}$} \\
\colhead{} & \colhead{(MJD)} & \colhead{(Hz)} & \colhead{($10^{-11}$ Hz s$^{-1}$)} & \colhead{(MJD)} & \colhead{($10^{-6}$ Hz)} & \colhead{($10^{-13}$ Hz s$^{-1}$)} & \colhead{($10^{-7}$)} & \colhead{($10^{-3}$)} & \colhead{($10^{42}$ erg)}
}

\startdata
1 & 59267.0 & 19.3526460(2)  & $-2.91964(7)$ & $59266.890(2)$ & $5.244(2)$  & $-0.80(2)$   & $2.710(1)$  & $2.75(5)$  & $4.007(2)$ \\ 
2 & 59867.0 & 19.35113496(5) & $-2.91800(1)$ & $59868.062(4)$ & $2.969(2)$  & $-1.44(1)$   & $1.5341(9)$ & $4.94(4)$  & $2.268(1)$ \\
3 & 60060.0 & 19.35064994(9) & $-2.92068(2)$ & $60063.47(3)$  & $0.375(2)$  & $-0.23(2)$   & $0.194(1)$  & $0.78(7)$  & $0.286(2)$ \\
4 & 61108.0 & 19.348009606(8)& $-2.939003(3)$& $61108.360(1)$ & $15.632(2)$ & $-3.08(2)$   & $8.080(1)$  & $10.48(6)$ & $11.940(2)$
\enddata

\tablecomments{Calculation of the change in rotational kinetic energy ($\Delta E_{\rm rot}$) uses a typically assumed moment of inertia of neutron stars $I = 10^{45}~{\rm g\ cm^2}$ \citep{handbook}. The quoted uncertainties in $\Delta E_{\rm rot}$ reflect only the measurement uncertainties in $\nu$ and $\Delta\nu$, not the systematic uncertainty in the assumed moment of inertia. }

\end{deluxetable*}

%% file: tb_glitch_parameters_rec.tex
\begin{deluxetable*}{cccc}

\tablecaption{Glitch Recovery Model Parameters} 
\tabletypesize{\footnotesize}
\label{tab:glitch-parameters-rec}

\setlength{\tabcolsep}{32pt}

\tablehead{\colhead{Parameter} & \colhead{Glitch \#1} & \colhead{Glitch \#2} & \colhead{Glitch \#3}} 

\startdata
Glitch Epoch (MJD)              & $59266.867(1)$             & $59867.941(6)$            & $60063.49(3)$  \\
$\Delta \nu$ (Hz)               & $1.321(1) \times 10^{-6}$  & $1.343(3) \times 10^{-6}$ & $1.43(9) \times 10^{-8}$ \\
$\Delta \dot\nu$ (Hz s$^{-1}$)  & $2.112(2) \times 10^{-14}$ & $3.1(2) \times 10^{-15}$  & $1.04(2) \times 10^{-14}$ \\
$\tau_{\rm d}^{(1)}$ (MJD)      & $136.46(8)$                & 76.5 \tiny{\it (fixed)}       & $94.8(4)$ \\
$\nu_{\rm d}^{(1)}$ (Hz)        & $4.965(1) \times 10^{-6}$  & $1.858(5) \times 10^{-6}$ & $3.334(8) \times 10^{-7}$ \\
$\tau_{\rm d}^{(2)}$ (MJD)      & $29.20(8)$                 & $11.1(2)$                 & ... \\
$\nu_{\rm d}^{(2)}$ (Hz)        & $-1.085(2) \times 10^{-6}$ & $-2.89(4) \times 10^{-7}$ & ... \\
\hline 
\enddata

\tablecomments{The pre-glitch $\nu$ and $\dot\nu$ are the same as those for glitch 1 in Table \ref{tab:glitch-parameters}. }

\end{deluxetable*}

%% file: tb_nustar_flux.tex
\begin{deluxetable}{ccc}

\tablecaption{NuSTAR Spectral Fitting Results} 
\tabletypesize{\footnotesize}
\label{tab:nustar_flux}

\tablehead{\colhead{Parameter} & \colhead{2020 September} & \colhead{2026 March}} 

\startdata
Model                                  & \multicolumn{2}{c}{\texttt{tbabs*pow}} \\
$N_{\rm H}$ ($10^{22}\ {\rm cm^{-2}}$)\tablenotemark{\tiny{a}} 
                                       & \multicolumn{2}{c}{0.89 \tiny{\it (fixed)}}  \\
$\Gamma_{\rm X}$                       & $1.47 \pm 0.06$         & $1.61 \pm 0.09$ \\
$\chi^2$(dof)                          & $47/47$                 & $18/21$ \\
$F_{\rm X}^{\rm abs}$ ($10^{-12}$~erg~s$^{-1}$~cm$^{-2}$)\tablenotemark{\tiny{b}} 
                                       & $4.7 \pm 0.2$ & $3.7_{-0.3}^{+0.2}$ \\
$F_{\rm X}^{\rm unabs}$ ($10^{-12}$~erg~s$^{-1}$~cm$^{-2}$)\tablenotemark{\tiny{c}} 
                                       & $1.13 \pm 0.03$ & $1.01 \pm 0.04$ \\
\hline 
\enddata

\tablenotetext{a}{The hydrogen column density, taken from \cite{pma+24} and fixed in the fit.}
\tablenotetext{b}{The 3--79~keV absorbed X-ray fluxes, taken from an average of FPMA and FPMB.}
\tablenotetext{c}{The 0.5--10~keV unabsorbed X-ray fluxes, calculated with the convolution \texttt{cflux} model.}

\tablecomments{
     All quoted uncertainties correspond to 1$\,\sigma$ confidence intervals. 
}

\end{deluxetable}

%% file: tb_det_summary.tex
\begin{deluxetable*}{cccccc}

\setlength{\tabcolsep}{22pt}

\tablecaption{Summary of X-ray Follow-up of RPP Glitches}
\tabletypesize{\footnotesize}

\label{tab:det_summary}

\tablehead{\colhead{PSR} & \colhead{$B_{\rm dip}$\tablenotemark{\tiny{a}} } & \colhead{$\Delta \nu/\nu$} & \colhead{$\Delta E_{\rm rot}$\tablenotemark{\tiny{b}} } & \colhead{$\Delta F_{\rm X}/F_{\rm X}$\tablenotemark{\tiny{c}} } & \colhead{Reference} \\
\colhead{...} & \colhead{(G)} & \colhead{...} & \colhead{(erg)} & \colhead{...} & \colhead{...} } 

\startdata
B0531$+$21 & $3.8 \times 10^{12}$ & $4.8 \times 10^{-7}$ & $1.7 \times 10^{43}$ & ... & 1, 2, 3 \\
B0833$-$45 & $3.4 \times 10^{12}$ & $3.1 \times 10^{-6}$ & $1.5 \times 10^{43}$ & ... & 4 \\
J1119$-$6127 & $4.1 \times 10^{13}$ & $5.8 \times 10^{-6}$ & $1.4 \times 10^{42}$ & $160$ & 5, 6 \\
J1846$-$0258  & $4.9 \times 10^{13} $ & $6.2 \times 10^{-6}$ & $2.3 \times 10^{42}$ & $7.7$ & 7, 8, 9 \\
{J2229$+$6114} & ${2.0 \times 10^{12}}$ & ${8.1 \times 10^{-7}}$ & ${1.2 \times 10^{43}}$ & ${<0.03}$ & 10, this work \\
\enddata

\tablenotetext{a}{The inferred surface $B$-field, taken from the ATNF pulsar catalog \citep{psrcat}.}
\tablenotetext{b}{The change in rotational kinetic energy due to the glitch, computed as $\Delta E_{\rm rot}\approx 4\pi^2 I \nu \Delta\nu$, assuming a canonical moment of inertia $I = 10^{45}~{\rm g\,cm^2}$ for all sources and that the neutron star rotates uniformly as a rigid body. }
\tablenotetext{c}{The fractional enhancement in 0.5--10\,keV unabsorbed X-ray fluxes, with $3\,\sigma$ upper limits for non-detections and ``...'' where no constraint was placed. }

\tablecomments{
     Parameters corresponding to the largest glitch from each source for which X-ray follow-up observations have been made. 
}

\tablerefs{(1) \citealt{ljg+15}; (2) \citealt{sls+18}; (3) \citealt{crab}; (4) \citealt{vela}; (5) \citealt{J1119}; (6) \citealt{wje+11}; (7) \citealt{lnk+11}; (8) \citealt{J1846_glitch_size}; (9) \citealt{J1846}; (10) \citealt{saa+23}. }

\end{deluxetable*}

%% file: refs.bib
@ARTICLE{chime_overview,
       author = {{CHIME Collaboration} and {Amiri}, Mandana and {Bandura}, Kevin and {Boskovic}, Anja and {Chen}, Tianyue and {Cliche}, Jean-Fran{\c{c}}ois and {Deng}, Meiling and {Denman}, Nolan and {Dobbs}, Matt and {Fandino}, Mateus and {Foreman}, Simon and {Halpern}, Mark and {Hanna}, David and {Hill}, Alex S. and {Hinshaw}, Gary and {H{\"o}fer}, Carolin and {Kania}, Joseph and {Klages}, Peter and {Landecker}, T.~L. and {MacEachern}, Joshua and {Masui}, Kiyoshi and {Mena-Parra}, Juan and {Milutinovic}, Nikola and {Mirhosseini}, Arash and {Newburgh}, Laura and {Nitsche}, Rick and {Ordog}, Anna and {Pen}, Ue-Li and {Pinsonneault-Marotte}, Tristan and {Polzin}, Ava and {Reda}, Alex and {Renard}, Andre and {Shaw}, J. Richard and {Siegel}, Seth R. and {Singh}, Saurabh and {Smegal}, Rick and {Tretyakov}, Ian and {van Gassen}, Kwinten and {Vanderlinde}, Keith and {Wang}, Haochen and {Wiebe}, Donald V. and {Willis}, James S. and {Wulf}, Dallas},
        title = "{An Overview of CHIME, the Canadian Hydrogen Intensity Mapping Experiment}",
      journal = {\apjs},
         year = 2022,
        month = aug,
       volume = {261},
       number = {2},
          eid = {29},
        pages = {29},
          doi = {10.3847/1538-4365/ac6fd9},
archivePrefix = {arXiv},
       eprint = {2201.07869},
 primaryClass = {astro-ph.IM},
       adsurl = {https://ui.adsabs.harvard.edu/abs/2022ApJS..261...29C}
}

@ARTICLE{chimepsr_overview,
       author = {{CHIME/Pulsar Collaboration} and {Amiri}, M. and {Bandura}, K.~M. and {Boyle}, P.~J. and {Brar}, C. and {Cliche}, J. -F. and {Crowter}, K. and {Cubranic}, D. and {Demorest}, P.~B. and {Denman}, N.~T. and {Dobbs}, M. and {Dong}, F.~Q. and {Fandino}, M. and {Fonseca}, E. and {Good}, D.~C. and {Halpern}, M. and {Hill}, A.~S. and {H{\"o}fer}, C. and {Kaspi}, V.~M. and {Landecker}, T.~L. and {Leung}, C. and {Lin}, H. -H. and {Luo}, J. and {Masui}, K.~W. and {McKee}, J.~W. and {Mena-Parra}, J. and {Meyers}, B.~W. and {Michilli}, D. and {Naidu}, A. and {Newburgh}, L. and {Ng}, C. and {Patel}, C. and {Pinsonneault-Marotte}, T. and {Ransom}, S.~M. and {Renard}, A. and {Scholz}, P. and {Shaw}, J.~R. and {Sikora}, A.~E. and {Stairs}, I.~H. and {Tan}, C.~M. and {Tendulkar}, S.~P. and {Tretyakov}, I. and {Vanderlinde}, K. and {Wang}, H. and {Wang}, X.},
        title = "{The CHIME Pulsar Project: System Overview}",
      journal = {\apjs},
         year = 2021,
        month = jul,
       volume = {255},
       number = {1},
          eid = {5},
        pages = {5},
          doi = {10.3847/1538-4365/abfdcb},
archivePrefix = {arXiv},
       eprint = {2008.05681},
 primaryClass = {astro-ph.IM},
       adsurl = {https://ui.adsabs.harvard.edu/abs/2021ApJS..255....5C}
}

@ARTICLE{clfd,
       author = {{Morello}, V. and {Barr}, E.~D. and {Cooper}, S. and {Bailes}, M. and {Bates}, S. and {Bhat}, N.~D.~R. and {Burgay}, M. and {Burke-Spolaor}, S. and {Cameron}, A.~D. and {Champion}, D.~J. and {Eatough}, R.~P. and {Flynn}, C.~M.~L. and {Jameson}, A. and {Johnston}, S. and {Keith}, M.~J. and {Keane}, E.~F. and {Kramer}, M. and {Levin}, L. and {Ng}, C. and {Petroff}, E. and {Possenti}, A. and {Stappers}, B.~W. and {van Straten}, W. and {Tiburzi}, C.},
        title = "{The High Time Resolution Universe survey - XIV. Discovery of 23 pulsars through GPU-accelerated reprocessing}",
      journal = {\mnras},
         year = 2019,
        month = mar,
       volume = {483},
       number = {3},
        pages = {3673-3685},
          doi = {10.1093/mnras/sty3328},
archivePrefix = {arXiv},
       eprint = {1811.04929},
 primaryClass = {astro-ph.IM},
       adsurl = {https://ui.adsabs.harvard.edu/abs/2019MNRAS.483.3673M}
}

@ARTICLE{psrchive,
       author = {{Hotan}, A.~W. and {van Straten}, W. and {Manchester}, R.~N.},
        title = "{PSRCHIVE and PSRFITS: An Open Approach to Radio Pulsar Data Storage and Analysis}",
      journal = {\pasa},
         year = 2004,
        month = jan,
       volume = {21},
       number = {3},
        pages = {302-309},
          doi = {10.1071/AS04022},
archivePrefix = {arXiv},
       eprint = {astro-ph/0404549},
 primaryClass = {astro-ph},
       adsurl = {https://ui.adsabs.harvard.edu/abs/2004PASA...21..302H}
}

@ARTICLE{J2229_discovery,
       author = {{Halpern}, J.~P. and {Camilo}, F. and {Gotthelf}, E.~V. and {Helfand}, D.~J. and {Kramer}, M. and {Lyne}, A.~G. and {Leighly}, K.~M. and {Eracleous}, M.},
        title = "{PSR J2229+6114: Discovery of an Energetic Young Pulsar in the Error Box of the EGRET Source 3EG J2227+6122}",
      journal = {\apjl},
         year = 2001,
        month = may,
       volume = {552},
       number = {2},
        pages = {L125-L128},
          doi = {10.1086/320347},
archivePrefix = {arXiv},
       eprint = {astro-ph/0104109},
 primaryClass = {astro-ph},
       adsurl = {https://ui.adsabs.harvard.edu/abs/2001ApJ...552L.125H}
}

@ARTICLE{bsa+22,
       author = {{Basu}, A. and {Shaw}, B. and {Antonopoulou}, D. and {Keith}, M.~J. and {Lyne}, A.~G. and {Mickaliger}, M.~B. and {Stappers}, B.~W. and {Weltevrede}, P. and {Jordan}, C.~A.},
        title = "{The Jodrell bank glitch catalogue: 106 new rotational glitches in 70 pulsars}",
      journal = {\mnras},
         year = 2022,
        month = mar,
       volume = {510},
       number = {3},
        pages = {4049-4062},
          doi = {10.1093/mnras/stab3336},
archivePrefix = {arXiv},
       eprint = {2111.06835},
 primaryClass = {astro-ph.HE},
       adsurl = {https://ui.adsabs.harvard.edu/abs/2022MNRAS.510.4049B}
}

@ARTICLE{ggy+22,
       author = {{G{\"u}gercino{\u{g}}lu}, E. and {Ge}, M.~Y. and {Yuan}, J.~P. and {Zhou}, S.~Q.},
        title = "{Glitches in four gamma-ray pulsars and inferences on the neutron star structure}",
      journal = {\mnras},
         year = 2022,
        month = mar,
       volume = {511},
       number = {1},
        pages = {425-439},
          doi = {10.1093/mnras/stac026},
archivePrefix = {arXiv},
       eprint = {2011.14788},
 primaryClass = {astro-ph.HE},
       adsurl = {https://ui.adsabs.harvard.edu/abs/2022MNRAS.511..425G}
}

@INPROCEEDINGS{arnaud96,
       author = {{Arnaud}, K.~A.},
        title = "{XSPEC: The First Ten Years}",
    booktitle = {Astronomical Data Analysis Software and Systems V},
         year = 1996,
       editor = {{Jacoby}, George H. and {Barnes}, Jeannette},
       series = {Astronomical Society of the Pacific Conference Series},
       volume = {101},
        month = jan,
        pages = {17},
       adsurl = {https://ui.adsabs.harvard.edu/abs/1996ASPC..101...17A}
}

@ARTICLE{pint,
       author = {{Luo}, Jing and {Ransom}, Scott and {Demorest}, Paul and {Ray}, Paul S. and {Archibald}, Anne and {Kerr}, Matthew and {Jennings}, Ross J. and {Bachetti}, Matteo and {van Haasteren}, Rutger and {Champagne}, Chloe A. and {Colen}, Jonathan and {Phillips}, Camryn and {Zimmerman}, Josef and {Stovall}, Kevin and {Lam}, Michael T. and {Jenet}, Fredrick A.},
        title = "{PINT: A Modern Software Package for Pulsar Timing}",
      journal = {\apj},
         year = 2021,
        month = apr,
       volume = {911},
       number = {1},
          eid = {45},
        pages = {45},
          doi = {10.3847/1538-4357/abe62f},
archivePrefix = {arXiv},
       eprint = {2012.00074},
 primaryClass = {astro-ph.IM},
       adsurl = {https://ui.adsabs.harvard.edu/abs/2021ApJ...911...45L}
}

@ARTICLE{champss_overview,
       author = {{Andrade}, Christopher and {Boyle}, P.~J. and {Brar}, Charanjot and {Cassity}, Alyssa and {Crowter}, Kathryn and {Cubranic}, Davor and {Denney}, Abigail K. and {Dong}, Fengqiu Adam and {Fonseca}, Emmanuel and {Kaspi}, Victoria M. and {Kumar}, Ajay and {K{\"u}nkel}, Lars and {L'Argent}, Magnus and {Lang}, Dustin and {Main}, Robert A. and {Masui}, Kiyoshi W. and {Mate}, Sujay and {Mena-Parra}, Juan and {Meyers}, Bradley W. and {Ng}, Cherry and {Pearlman}, Aaron B. and {Pen}, Ue-Li and {Ransom}, Scott M. and {Roman}, Alexander P. and {Smith}, Kendrick and {Squillace}, Reynier and {Stairs}, Ingrid and {Tan}, Chia Min and {Tarabout}, Laurent and {Xia}, Wenke and {Zegmott}, Tarik J. and {The Champss Collaboration}},
        title = "{CHIME All-sky Multiday Pulsar Stacking Search (CHAMPSS): System Overview and First Discoveries}",
      journal = {\apj},
         year = 2025,
        month = sep,
       volume = {990},
       number = {1},
          eid = {50},
        pages = {50},
          doi = {10.3847/1538-4357/adeb51},
archivePrefix = {arXiv},
       eprint = {2504.16293},
 primaryClass = {astro-ph.HE},
       adsurl = {https://ui.adsabs.harvard.edu/abs/2025ApJ...990...50A}
}

@ARTICLE{J1119,
       author = {{Archibald}, R.~F. and {Kaspi}, V.~M. and {Tendulkar}, S.~P. and {Scholz}, P.},
        title = "{A Magnetar-like Outburst from a High-B Radio Pulsar}",
      journal = {\apjl},
         year = 2016,
        month = sep,
       volume = {829},
       number = {1},
          eid = {L21},
        pages = {L21},
          doi = {10.3847/2041-8205/829/1/L21},
archivePrefix = {arXiv},
       eprint = {1608.01007},
 primaryClass = {astro-ph.HE},
       adsurl = {https://ui.adsabs.harvard.edu/abs/2016ApJ...829L..21A}
}

@ARTICLE{J1846,
       author = {{Gavriil}, F.~P. and {Gonzalez}, M.~E. and {Gotthelf}, E.~V. and {Kaspi}, V.~M. and {Livingstone}, M.~A. and {Woods}, P.~M.},
        title = "{Magnetar-Like Emission from the Young Pulsar in Kes 75}",
      journal = {Science},
         year = 2008,
        month = mar,
       volume = {319},
       number = {5871},
        pages = {1802},
          doi = {10.1126/science.1153465},
archivePrefix = {arXiv},
       eprint = {0802.1704},
 primaryClass = {astro-ph},
       adsurl = {https://ui.adsabs.harvard.edu/abs/2008Sci...319.1802G}
}

@ARTICLE{vela,
       author = {{Dodson}, R.~G. and {McCulloch}, P.~M. and {Lewis}, D.~R.},
        title = "{High Time Resolution Observations of the January 2000 Glitch in the Vela Pulsar}",
      journal = {\apjl},
         year = 2002,
        month = jan,
       volume = {564},
       number = {2},
        pages = {L85-L88},
          doi = {10.1086/339068},
archivePrefix = {arXiv},
       eprint = {astro-ph/0201005},
 primaryClass = {astro-ph},
       adsurl = {https://ui.adsabs.harvard.edu/abs/2002ApJ...564L..85D}
}

@ARTICLE{crab,
       author = {{Zhang}, Xinyuan and {Shuai}, Ping and {Huang}, Liangwei and {Chen}, Shaolong and {Du}, Yuanjie},
        title = "{X-Ray Observation of the 2017 November Glitch in the Crab Pulsar}",
      journal = {\apj},
         year = 2018,
        month = oct,
       volume = {866},
       number = {2},
          eid = {82},
        pages = {82},
          doi = {10.3847/1538-4357/aade46},
       adsurl = {https://ui.adsabs.harvard.edu/abs/2018ApJ...866...82Z}
}

@ARTICLE{chimefrb-overview,
       author = {{CHIME/FRB Collaboration} and {Amiri}, M. and {Bandura}, K. and {Berger}, P. and {Bhardwaj}, M. and {Boyce}, M.~M. and {Boyle}, P.~J. and {Brar}, C. and {Burhanpurkar}, M. and {Chawla}, P. and {Chowdhury}, J. and {Cliche}, J.-F. and {Cranmer}, M.~D. and {Cubranic}, D. and {Deng}, M. and {Denman}, N. and {Dobbs}, M. and {Fandino}, M. and {Fonseca}, E. and {Gaensler}, B.~M. and {Giri}, U. and {Gilbert}, A.~J. and {Good}, D.~C. and {Guliani}, S. and {Halpern}, M. and {Hinshaw}, G. and {H{\"o}fer}, C. and {Josephy}, A. and {Kaspi}, V.~M. and {Landecker}, T.~L. and {Lang}, D. and {Liao}, H. and {Masui}, K.~W. and {Mena-Parra}, J. and {Naidu}, A. and {Newburgh}, L.~B. and {Ng}, C. and {Patel}, C. and {Pen}, U.-L. and {Pinsonneault-Marotte}, T. and {Pleunis}, Z. and {Rafiei Ravandi}, M. and {Ransom}, S.~M. and {Renard}, A. and {Scholz}, P. and {Sigurdson}, K. and {Siegel}, S.~R. and {Smith}, K.~M. and {Stairs}, I.~H. and {Tendulkar}, S.~P. and {Vanderlinde}, K. and {Wiebe}, D.~V.},
        title = "{The CHIME Fast Radio Burst Project: System Overview}",
      journal = {\apj},
         year = 2018,
        month = aug,
       volume = {863},
       number = {1},
          eid = {48},
        pages = {48},
          doi = {10.3847/1538-4357/aad188},
archivePrefix = {arXiv},
       eprint = {1803.11235},
 primaryClass = {astro-ph.IM},
       adsurl = {https://ui.adsabs.harvard.edu/abs/2018ApJ...863...48C}
}

@BOOK{handbook,
       author = {{Lorimer}, D.~R. and {Kramer}, M.},
        title = "{Handbook of Pulsar Astronomy}",
         year = 2004,
       volume = {4},
       adsurl = {https://ui.adsabs.harvard.edu/abs/2004hpa..book.....L}
}

@ARTICLE{psrcat,
       author = {{Manchester}, R.~N. and {Hobbs}, G.~B. and {Teoh}, A. and {Hobbs}, M.},
        title = "{The Australia Telescope National Facility Pulsar Catalogue}",
      journal = {\aj},
         year = 2005,
        month = apr,
       volume = {129},
       number = {4},
        pages = {1993-2006},
          doi = {10.1086/428488},
archivePrefix = {arXiv},
       eprint = {astro-ph/0412641},
 primaryClass = {astro-ph},
       adsurl = {https://ui.adsabs.harvard.edu/abs/2005AJ....129.1993M}
}

@ARTICLE{J1846_glitch_size,
       author = {{Livingstone}, Margaret A. and {Kaspi}, Victoria M. and {Gavriil}, Fotis P. and {Manchester}, Richard N. and {Gotthelf}, E.~V.~G. and {Kuiper}, Lucien},
        title = "{New phase-coherent measurements of pulsar braking indices}",
      journal = {\apss},
         year = 2007,
        month = apr,
       volume = {308},
       number = {1-4},
        pages = {317-323},
          doi = {10.1007/s10509-007-9320-3},
archivePrefix = {arXiv},
       eprint = {astro-ph/0702196},
 primaryClass = {astro-ph},
       adsurl = {https://ui.adsabs.harvard.edu/abs/2007Ap&SS.308..317L}
}

@ARTICLE{pp+11,
       author = {{Perna}, Rosalba and {Pons}, Jose A.},
        title = "{A Unified Model of the Magnetar and Radio Pulsar Bursting Phenomenology}",
      journal = {\apjl},
         year = 2011,
        month = feb,
       volume = {727},
       number = {2},
          eid = {L51},
        pages = {L51},
          doi = {10.1088/2041-8205/727/2/L51},
archivePrefix = {arXiv},
       eprint = {1101.1098},
 primaryClass = {astro-ph.HE},
       adsurl = {https://ui.adsabs.harvard.edu/abs/2011ApJ...727L..51P}
}

@ARTICLE{kgw+03,
       author = {{Kaspi}, V.~M. and {Gavriil}, F.~P. and {Woods}, P.~M. and {Jensen}, J.~B. and {Roberts}, M.~S.~E. and {Chakrabarty}, D.},
        title = "{A Major Soft Gamma Repeater-like Outburst and Rotation Glitch in the No-longer-so-anomalous X-Ray Pulsar 1E 2259+586}",
      journal = {\apjl},
         year = 2003,
        month = may,
       volume = {588},
       number = {2},
        pages = {L93-L96},
          doi = {10.1086/375683},
archivePrefix = {arXiv},
       eprint = {astro-ph/0304205},
 primaryClass = {astro-ph},
       adsurl = {https://ui.adsabs.harvard.edu/abs/2003ApJ...588L..93K}
}

@ARTICLE{kas03,
       author = {{Kaspi}, Victoria M.},
        title = "{Grand unification of neutron stars}",
      journal = {Proceedings of the National Academy of Science},
         year = 2010,
        month = apr,
       volume = {107},
       number = {16},
        pages = {7147-7152},
          doi = {10.1073/pnas.1000812107},
archivePrefix = {arXiv},
       eprint = {1005.0876},
 primaryClass = {astro-ph.HE},
       adsurl = {https://ui.adsabs.harvard.edu/abs/2010PNAS..107.7147K}
}

@ARTICLE{J1846_2020,
       author = {{Hu}, Chin-Ping and {Kuiper}, Lucien and {Harding}, Alice K. and {Younes}, George and {Blumer}, Harsha and {Ho}, Wynn C.~G. and {Enoto}, Teruaki and {Espinoza}, Crist{\'o}bal M. and {Gendreau}, Keith},
        title = "{A NICER View on the 2020 Magnetar-like Outburst of PSR J1846-0258}",
      journal = {\apj},
         year = 2023,
        month = aug,
       volume = {952},
       number = {2},
          eid = {120},
        pages = {120},
          doi = {10.3847/1538-4357/acd850},
archivePrefix = {arXiv},
       eprint = {2306.00902},
 primaryClass = {astro-ph.HE},
       adsurl = {https://ui.adsabs.harvard.edu/abs/2023ApJ...952..120H}
}

@ARTICLE{wje+11,
       author = {{Weltevrede}, Patrick and {Johnston}, Simon and {Espinoza}, Crist{\'o}bal M.},
        title = "{The glitch-induced identity changes of PSR J1119-6127}",
      journal = {\mnras},
         year = 2011,
        month = mar,
       volume = {411},
       number = {3},
        pages = {1917-1934},
          doi = {10.1111/j.1365-2966.2010.17821.x},
archivePrefix = {arXiv},
       eprint = {1010.0857},
 primaryClass = {astro-ph.SR},
       adsurl = {https://ui.adsabs.harvard.edu/abs/2011MNRAS.411.1917W}
}

@article{saa+23,
  author = {{Smith}, D. A. and {Abdollahi}, S. and {Ajello}, M. and {Bailes}, M. and {Baldini}, L. and {Ballet}, J. and {Baring}, M. G. and {Bassa}, C. and {Becerra Gonzalez}, J. and {Bellazzini}, R. and {Berretta}, A. and {Bhattacharyya}, B. and {Bissaldi}, E. and {et al.}},
  title = {The Third Fermi Large Area Telescope Catalog of Gamma-Ray Pulsars},
  journal = {apj},
  year = {2023},
  month = {December},
  volume = {958},
  number = {2},
  pages = {191},
  doi = {10.3847/1538-4357/acee67},
  url = {https://ui.adsabs.harvard.edu/abs/2023ApJ...958..191S}
}

@ARTICLE{apt,
       author = {{Phillips}, Camryn and {Ransom}, Scott},
        title = "{Algorithmic Pulsar Timing}",
      journal = {\aj},
         year = 2022,
        month = feb,
       volume = {163},
       number = {2},
          eid = {84},
        pages = {84},
          doi = {10.3847/1538-3881/ac403e},
       adsurl = {https://ui.adsabs.harvard.edu/abs/2022AJ....163...84P}
}

@ARTICLE{sls+18,
       author = {{Shaw}, B. and {Lyne}, A.~G. and {Stappers}, B.~W. and {Weltevrede}, P. and {Bassa}, C.~G. and {Lien}, A.~Y. and {Mickaliger}, M.~B. and {Breton}, R.~P. and {Jordan}, C.~A. and {Keith}, M.~J. and {Krimm}, H.~A.},
        title = "{The largest glitch observed in the Crab pulsar}",
      journal = {\mnras},
         year = 2018,
        month = aug,
       volume = {478},
       number = {3},
        pages = {3832-3840},
          doi = {10.1093/mnras/sty1294},
archivePrefix = {arXiv},
       eprint = {1805.05110},
 primaryClass = {astro-ph.HE},
       adsurl = {https://ui.adsabs.harvard.edu/abs/2018MNRAS.478.3832S}
}

@article{lnk+11,
  author = {{Livingstone}, M. A. and {Ng}, {C.-Y.} and {Kaspi}, V. M. and {Gavriil}, F. P. and {Gotthelf}, E. V.},
  title = {Post-outburst Observations of the Magnetically Active Pulsar J1846-0258. A New Braking Index, Increased Timing Noise, and Radiative Recovery},
  journal = {apj},
  year = {2011},
  month = {April},
  volume = {730},
  pages = {66-+},
  doi = {10.1088/0004-637X/730/2/66},
  url = {http://adsabs.harvard.edu/abs/2011ApJ...730...66L}
}

@article{ljg+15,
  author = {{Lyne}, A. G. and {Jordan}, C. A. and {Graham-Smith}, F. and {Espinoza}, C. M. and {Stappers}, B. W. and {Weltevrede}, P.},
  title = {45 years of rotation of the Crab pulsar},
  journal = {mnras},
  year = {2015},
  month = {January},
  volume = {446},
  pages = {857-864},
  doi = {10.1093/mnras/stu2118},
  url = {http://adsabs.harvard.edu/abs/2015MNRAS.446..857L}
}

@article{emcee,
   author = {{Foreman-Mackey}, D. and {Hogg}, D.~W. and {Lang}, D. and {Goodman}, J.},
    title = {emcee: The MCMC Hammer},
  journal = {PASP},
     year = 2013,
   volume = 125,
    pages = {306-312},
   eprint = {1202.3665},
      doi = {10.1086/670067}
}

@ARTICLE{good+21,
       author = {{Good}, D.~C. and {Andersen}, B.~C. and {Chawla}, P. and {Crowter}, K. and {Dong}, F.~Q. and {Fonseca}, E. and {Meyers}, B.~W. and {Ng}, C. and {Pleunis}, Z. and {Ransom}, S.~M. and {Stairs}, I.~H. and {Tan}, C.~M. and {Bhardwaj}, M. and {Boyle}, P.~J. and {Dobbs}, M. and {Gaensler}, B.~M. and {Kaspi}, V.~M. and {Masui}, K.~W. and {Naidu}, A. and {Rafiei-Ravandi}, M. and {Scholz}, P. and {Smith}, K.~M. and {Tendulkar}, S.~P.},
        title = "{First Discovery of New Pulsars and RRATs with CHIME/FRB}",
      journal = {\apj},
         year = 2021,
        month = nov,
       volume = {922},
       number = {1},
          eid = {43},
        pages = {43},
          doi = {10.3847/1538-4357/ac1da6},
archivePrefix = {arXiv},
       eprint = {2012.02320},
 primaryClass = {astro-ph.HE},
       adsurl = {https://ui.adsabs.harvard.edu/abs/2021ApJ...922...43G}
}

@ARTICLE{pma+24,
       author = {{Pope}, I. and {Mori}, K. and {Abdelmaguid}, M. and {Gelfand}, J.~D. and {Reynolds}, S.~P. and {Safi-Harb}, S. and {Hailey}, C.~J. and {An}, H. and {Bangale}, P. and {Batista}, P. and {Benbow}, W. and {Buckley}, J.~H. and {Capasso}, M. and {Christiansen}, J.~L. and {Chromey}, A.~J. and {Falcone}, A. and {Feng}, Q. and {Finley}, J.~P. and {Foote}, Juniper and {Gallagher}, G. and {Hanlon}, W.~F. and {Hanna}, D. and {Hervet}, O. and {Holder}, J. and {Humensky}, T.~B. and {Jin}, W. and {Kaaret}, P. and {Kertzman}, M. and {Kieda}, D. and {Kleiner}, T.~K. and {Korzoun}, N. and {Krennrich}, F. and {Kumar}, S. and {Lang}, M.~J. and {Maier}, G. and {McGrath}, C.~E. and {Mooney}, C.~L. and {Moriarty}, P. and {Mukherjee}, R. and {O'Brien}, S. and {Ong}, R.~A. and {Park}, N. and {Patel}, S.~R. and {Pfrang}, K. and {Pohl}, M. and {Pueschel}, E. and {Quinn}, J. and {Ragan}, K. and {Reynolds}, P.~T. and {Roache}, E. and {Sadeh}, I. and {Saha}, L. and {Sembroski}, G.~H. and {Tak}, D. and {Tucci}, J.~V. and {Weinstein}, A. and {Williams}, D.~A. and {Woo}, J. and {VERITAS Collaboration}},
        title = "{A Multiwavelength Investigation of PSR J2229+6114 and its Pulsar Wind Nebula in the Radio, X-Ray, and Gamma-Ray Bands}",
      journal = {\apj},
         year = 2024,
        month = jan,
       volume = {960},
       number = {1},
          eid = {75},
        pages = {75},
          doi = {10.3847/1538-4357/ad0120},
archivePrefix = {arXiv},
       eprint = {2310.04512},
 primaryClass = {astro-ph.HE},
       adsurl = {https://ui.adsabs.harvard.edu/abs/2024ApJ...960...75P}
}

@ARTICLE{ckl+00,
       author = {{Camilo}, F. and {Kaspi}, V.~M. and {Lyne}, A.~G. and {Manchester}, R.~N. and {Bell}, J.~F. and {D'Amico}, N. and {McKay}, N.~P.~F. and {Crawford}, F.},
        title = "{Discovery of Two High Magnetic Field Radio Pulsars}",
      journal = {\apj},
         year = 2000,
        month = sep,
       volume = {541},
       number = {1},
        pages = {367-373},
          doi = {10.1086/309435},
archivePrefix = {arXiv},
       eprint = {astro-ph/0004330},
 primaryClass = {astro-ph},
       adsurl = {https://ui.adsabs.harvard.edu/abs/2000ApJ...541..367C}
}

@ARTICLE{gvb+00,
       author = {{Gotthelf}, E.~V. and {Vasisht}, G. and {Boylan-Kolchin}, M. and {Torii}, K.},
        title = "{A 700 Year-old Pulsar in the Supernova Remnant Kesteven 75}",
      journal = {\apjl},
         year = 2000,
        month = oct,
       volume = {542},
       number = {1},
        pages = {L37-L40},
          doi = {10.1086/312923},
archivePrefix = {arXiv},
       eprint = {astro-ph/0008097},
 primaryClass = {astro-ph},
       adsurl = {https://ui.adsabs.harvard.edu/abs/2000ApJ...542L..37G}
}

@ARTICLE{bkk+18,
       author = {{Brook}, P.~R. and {Karastergiou}, A. and {McLaughlin}, M.~A. and {Lam}, M.~T. and {Arzoumanian}, Z. and {Chatterjee}, S. and {Cordes}, J.~M. and {Crowter}, K. and {DeCesar}, M. and {Demorest}, P.~B. and {Dolch}, T. and {Ellis}, J.~A. and {Ferdman}, R.~D. and {Ferrara}, E. and {Fonseca}, E. and {Gentile}, P.~A. and {Jones}, G. and {Jones}, M.~L. and {Lazio}, T.~J.~W. and {Levin}, L. and {Lorimer}, D.~R. and {Lynch}, R.~S. and {Ng}, C. and {Nice}, D.~J. and {Pennucci}, T.~T. and {Ransom}, S.~M. and {Ray}, P.~S. and {Spiewak}, R. and {Stairs}, I.~H. and {Stinebring}, D.~R. and {Stovall}, K. and {Swiggum}, J.~K. and {Zhu}, W.~W.},
        title = "{The NANOGrav 11-year Data Set: Pulse Profile Variability}",
      journal = {\apj},
         year = 2018,
        month = dec,
       volume = {868},
       number = {2},
          eid = {122},
        pages = {122},
          doi = {10.3847/1538-4357/aae9e3},
archivePrefix = {arXiv},
       eprint = {1810.08269},
 primaryClass = {astro-ph.HE},
       adsurl = {https://ui.adsabs.harvard.edu/abs/2018ApJ...868..122B}
}

@ARTICLE{jcc+24,
       author = {{Jennings}, Ross J. and {Cordes}, James M. and {Chatterjee}, Shami and {McLaughlin}, Maura A. and {Demorest}, Paul B. and {Arzoumanian}, Zaven and {Baker}, Paul T. and {Blumer}, Harsha and {Brook}, Paul R. and {Cohen}, Tyler and {Crawford}, Fronefield and {Cromartie}, H. Thankful and {DeCesar}, Megan E. and {Dolch}, Timothy and {Ferrara}, Elizabeth C. and {Fonseca}, Emmanuel and {Good}, Deborah C. and {Hazboun}, Jeffrey S. and {Jones}, Megan L. and {Kaplan}, David L. and {Lam}, Michael T. and {Lazio}, T. Joseph W. and {Lorimer}, Duncan R. and {Luo}, Jing and {Lynch}, Ryan S. and {McKee}, James W. and {Madison}, Dustin R. and {Meyers}, Bradley W. and {Mingarelli}, Chiara M.~F. and {Nice}, David J. and {Pennucci}, Timothy T. and {Perera}, Benetge B.~P. and {Pol}, Nihan S. and {Ransom}, Scott M. and {Ray}, Paul S. and {Shapiro-Albert}, Brent J. and {Siemens}, Xavier and {Stairs}, Ingrid H. and {Stinebring}, Daniel R. and {Swiggum}, Joseph K. and {Tan}, Chia Min and {Taylor}, Stephen R. and {Vigeland}, Sarah J. and {Witt}, Caitlin A.},
        title = "{An Unusual Pulse Shape Change Event in PSR J1713+0747 Observed with the Green Bank Telescope and CHIME}",
      journal = {\apj},
         year = 2024,
        month = apr,
       volume = {964},
       number = {2},
          eid = {179},
        pages = {179},
          doi = {10.3847/1538-4357/ad2930},
archivePrefix = {arXiv},
       eprint = {2210.12266},
 primaryClass = {astro-ph.HE},
       adsurl = {https://ui.adsabs.harvard.edu/abs/2024ApJ...964..179J}
}

@ARTICLE{chime_magnetar,
       author = {{CHIME/FRB Collaboration} and {Andersen}, B.~C. and {Bandura}, K.~M. and {Bhardwaj}, M. and {Bij}, A. and {Boyce}, M.~M. and {Boyle}, P.~J. and {Brar}, C. and {Cassanelli}, T. and {Chawla}, P. and {Chen}, T. and {Cliche}, J.-F. and {Cook}, A. and {Cubranic}, D. and {Curtin}, A.~P. and {Denman}, N.~T. and {Dobbs}, M. and {Dong}, F.~Q. and {Fandino}, M. and {Fonseca}, E. and {Gaensler}, B.~M. and {Giri}, U. and {Good}, D.~C. and {Halpern}, M. and {Hill}, A.~S. and {Hinshaw}, G.~F. and {H{\"o}fer}, C. and {Josephy}, A. and {Kania}, J.~W. and {Kaspi}, V.~M. and {Landecker}, T.~L. and {Leung}, C. and {Li}, D.~Z. and {Lin}, H.-H. and {Masui}, K.~W. and {McKinven}, R. and {Mena-Parra}, J. and {Merryfield}, M. and {Meyers}, B.~W. and {Michilli}, D. and {Milutinovic}, N. and {Mirhosseini}, A. and {M{\"u}nchmeyer}, M. and {Naidu}, A. and {Newburgh}, L.~B. and {Ng}, C. and {Patel}, C. and {Pen}, U.-L. and {Pinsonneault-Marotte}, T. and {Pleunis}, Z. and {Quine}, B.~M. and {Rafiei-Ravandi}, M. and {Rahman}, M. and {Ransom}, S.~M. and {Renard}, A. and {Sanghavi}, P. and {Scholz}, P. and {Shaw}, J.~R. and {Shin}, K. and {Siegel}, S.~R. and {Singh}, S. and {Smegal}, R.~J. and {Smith}, K.~M. and {Stairs}, I.~H. and {Tan}, C.~M. and {Tendulkar}, S.~P. and {Tretyakov}, I. and {Vanderlinde}, K. and {Wang}, H. and {Wulf}, D. and {Zwaniga}, A.~V.},
        title = "{A bright millisecond-duration radio burst from a Galactic magnetar}",
      journal = {\nat},
         year = 2020,
        month = nov,
       volume = {587},
       number = {7832},
        pages = {54-58},
          doi = {10.1038/s41586-020-2863-y},
archivePrefix = {arXiv},
       eprint = {2005.10324},
 primaryClass = {astro-ph.HE},
       adsurl = {https://ui.adsabs.harvard.edu/abs/2020Natur.587...54C}
}

@ARTICLE{zgy+22,
       author = {{Zhou}, Shiqi and {G{\"u}gercino{\u{g}}lu}, Erbil and {Yuan}, Jianping and {Ge}, Mingyu and {Yu}, Cong},
        title = "{Pulsar Glitches: A Review}",
      journal = {Universe},
         year = 2022,
        month = dec,
       volume = {8},
       number = {12},
          eid = {641},
        pages = {641},
          doi = {10.3390/universe8120641},
archivePrefix = {arXiv},
       eprint = {2211.13885},
 primaryClass = {astro-ph.HE},
       adsurl = {https://ui.adsabs.harvard.edu/abs/2022Univ....8..641Z}
}

@ARTICLE{kb17,
       author = {{Kaspi}, Victoria M. and {Beloborodov}, Andrei M.},
        title = "{Magnetars}",
      journal = {\araa},
         year = 2017,
        month = aug,
       volume = {55},
       number = {1},
        pages = {261-301},
          doi = {10.1146/annurev-astro-081915-023329},
archivePrefix = {arXiv},
       eprint = {1703.00068},
 primaryClass = {astro-ph.HE},
       adsurl = {https://ui.adsabs.harvard.edu/abs/2017ARA&A..55..261K}
}
